\documentclass[]{llncs}

\usepackage{pslatex}
\usepackage{amsmath}
\usepackage{leftidx}
\usepackage{epsfig}
\usepackage{paralist}
\usepackage{graphics}
\usepackage{stmaryrd}
\usepackage{txfonts}
\usepackage{framed}
\usepackage{makecell}
\usepackage{url}
\usepackage[colorlinks]{hyperref}
\usepackage{subfig}
\usepackage{wrapfig}

\usepackage{algorithm}
\usepackage{algorithmicx}
\usepackage[noend]{algpseudocode}

\usepackage[draft]{commenting}

\declareauthor{ri}{Radu}{blue}
\declareauthor{mb}{Marius}{red}
\declareauthor{js}{Joseph}{magenta}

\newtheorem{fact}{Fact}
\newcommand{\np}{$\mathsf{NP}$}

\newcommand{\rbr}{{\bf ]\!]}}
\newcommand{\lbr}{{\bf [\![}}
\newcommand{\sem}[1]{\lbr #1 \rbr}

\newcommand{\set}[1]{\{ #1 \}}
\newcommand{\tuple}[1]{\langle #1 \rangle}
\renewcommand{\vec}[1]{\mathbf #1}

\newcommand{\isdef}{\stackrel{\scriptscriptstyle{\mathsf{def}}}{=}}

\newcommand{\len}[1]{{\left|{#1}\right|}}
\newcommand{\card}[1]{{||{#1}||}}

\newcommand{\arrow}[2]{\xrightarrow{{\scriptscriptstyle #1}}_{{\scriptscriptstyle #2}}}

\newcommand{\nat}{{\bf \mathbb{N}}}
\newcommand{\zed}{{\bf \mathbb{Z}}}

\renewcommand{\paragraph}[1]{\noindent{\bf #1}}

\newcommand{\I}{\mathcal{I}}
\newcommand{\U}{\mathfrak{U}}
\newcommand{\true}{\top}
\newcommand{\false}{\bot}

\newcommand{\vars}{\mathsf{Var}}
\newcommand{\bvars}{\mathsf{BVar}}
\newcommand{\preds}{\mathsf{Pred}}

\newcommand{\pos}[1]{{#1}^+}
\newcommand{\ppos}[1]{{#1}^\oplus}
\newcommand{\voc}[1]{\mathrm{P}({#1})}
\newcommand{\posvoc}[1]{\mathrm{P}^+({#1})}
\newcommand{\negvoc}[1]{\mathrm{P}^-({#1})}
\newcommand{\poslan}[1]{\mathcal{L}^+({#1})}
\newcommand{\neglan}[1]{\mathcal{L}^-({#1})}

\newcommand{\anet}{N}
\newcommand{\amarkednet}{\mathcal{N}}
\newcommand{\places}{S}
\newcommand{\trans}{T}
\newcommand{\atrans}{\mathfrak{t}}
\newcommand{\edges}{E}
\newcommand{\pre}[1]{\leftidx{^\bullet}{\text{${#1}$}}}
\newcommand{\post}[1]{{#1}^\bullet}
\newcommand{\amark}{\mathrm{m}}
\newcommand{\reach}[1]{\mathcal{R}({#1})}

\newcommand{\dead}[1]{\mathcal{D}({#1})}

\newcommand{\alltrap}[1]{\mathit{Trap}({#1})}

\newcommand{\alldead}[1]{\mathit{Dead}({#1})}
\newcommand{\init}[1]{\mathit{Init}({#1})}

\newcommand{\acomptype}{\mathcal{C}}

\newcommand{\maxnf}{\mathsf{M}}
\newcommand{\maxn}[1]{\maxnf({#1})}

\newcommand{\typeno}[2]{{#1}^{\scriptscriptstyle{{#2}}}}
\newcommand{\ports}{\mathsf{P}}
\newcommand{\states}{\mathsf{S}}
\newcommand{\initstate}{{s_0}}
\newcommand{\rules}{\Delta}
\newcommand{\asys}{\mathcal{S}}
\newcommand{\interform}{\Gamma}
\newcommand{\trapconstraint}[1]{\Theta({#1})}
\newcommand{\deadconstraint}[1]{\Delta({#1})}

\newcommand{\unfolding}[1]{\mathcal{U}({#1})}

\newcommand{\mil}{$\mathsf{MIL}$}
\newcommand{\apred}{\mathsf{pred}}
\newcommand{\qelim}[1]{\mathrm{qe}({#1})}

\newcommand{\dualnopar}[1]{{#1}^{\sim}}
\newcommand{\dual}[1]{\left({#1}\right)^{\sim}}
\newcommand{\upclose}[1]{{#1}\!\!\uparrow}

\newcommand{\distinct}{\mathrm{distinct}}

\newcommand{\minsem}[1]{\sem{#1}^{\mathrm{\mu}}}
\newcommand{\minequiv}{\equiv^{\mathrm{\mu}}}

\newcommand{\bool}[2]{\mathrm{B}_{{#1}}\left({#2}\right)}
\newcommand{\closure}[1]{{#1}\raisebox{1pt}{$\uparrow$}}
\newcommand{\compt}[2]{t_{#1}^{\scriptscriptstyle{#2}}}
\newcommand{\setord}{\unlhd^\dagger}
\newcommand{\setordneq}{\lhd^\dagger}
\renewcommand{\proof}[1]{\ifLongVersion \noindent\emph{Proof}: {#1} \vspace*{\baselineskip}\else\fi}

\usepackage{setspace}

\newif\ifLongVersion\LongVersiontrue

\begin{document}
%%%%%%%%%%%%%%%%%%%%%%%%%%%%%%%%%%%%%%%%%%%%%%%%%%%%%%%%%%%%%%%%%%%%%%%%%%%%%%%

\setlength{\belowdisplayskip}{2pt} \setlength{\belowdisplayshortskip}{1pt}
\setlength{\abovedisplayskip}{2pt} \setlength{\abovedisplayshortskip}{1pt}

\title{Checking Deadlock-Freedom of Parametric Component-Based Systems}

\author{Marius Bozga, Radu Iosif and Joseph Sifakis}
\institute{Univ. Grenoble Alpes, CNRS, %
  Grenoble INP\footnote{Institute of Engineering Univ. Grenoble Alpes}, VERIMAG, 38000 
   Grenoble France 
   \url{{Marius.Bozga,Radu.Iosif,Joseph.Sifakis}@univ-grenoble-alpes.fr}  
}

\maketitle
\begin{abstract}
We propose an automated method for computing inductive invariants used
to proving deadlock freedom of parametric component-based systems. The
method generalizes the approach for computing structural trap
invariants from bounded to parametric systems with general
architectures. It symbolically extracts trap invariants from
interaction formulae defining the system architecture. The paper
presents the theoretical foundations of the method, including new
results for the first order monadic logic and proves its soundness. It
also reports on a preliminary experimental evaluation on several
textbook examples.
\end{abstract}

Modern computing systems exhibit dynamic and reconfigurable behavior.
To tackle the complexity of such systems, engineers extensively use
architectures that enforce, by construction, essential properties,
such as fault tolerance or mutual exclusion. Architectures can be
viewed as parametric operators that take as arguments instances of
components of given types and enforce a characteristic property. For
instance, client-server architectures enforce atomicity and resilience
of transactions, for any numbers of clients and servers. Similarly,
token-ring architectures enforce mutual exclusion between any number
of components in the ring.

Parametric verification is an extremely relevant and challenging
problem in systems engineering. In contrast to the verification of
bounded systems, consisting of a known set of components, there exist
no general methods and tools succesfully applied to parametric
systems. Verification problems for very simple parametric systems,
even with finite-state components, are typically intractable
\cite{GermanS92,Bloem2015}. Most work in this area puts emphasis on
limitations determined mainly by three
criteria \begin{inparaenum}[(1)]
\item the topology of the architecture, 
\item the coordination primitives, and 
\item the properties to be verified. 
\end{inparaenum}

The main decidability results reduce parametric verification to the
verification of a bounded number of instances of finite state
components. Several methods try to determine a cut-off size of the
system, i.e.\ the minimal size for which if a property holds, then it
holds for any size, e.g.\ Suzuki \cite{Suzuki88}, Emerson and Namjoshi
\cite{EmersonN95}. Other methods
identify systems with well-structured transition relations, for which
symbolic enumeration of reachable states is feasible \cite{Abdulla10}
or reduce to known decidable problems, such as reachability in vector
addition systems \cite{GermanS92}. Typically, these methods apply to
systems with global coordination.  When theoretical decidability is
not of concern, semi-algorithmic techniques such as \emph{regular
model checking}
\cite{KestenMalerMarcusPnueliShahar01,AbdullaHendaDelzannoRezine07},
SMT-based \emph{bounded model checking}
\cite{AlbertiGhilardiSharygina14,ConchonGoelKrsticMebsoutZaidi12},
\emph{abstraction}
\cite{BaukusBensalemLakhnechStahl00,BouajjaniHabermehlVojnar04} and
\emph{automata learning} \cite{ChenHongLinRummer17} can be used to
deal with more general classes of The interested reader can find a
complete survey on parameterized model checking by Bloem et
al. \cite{Bloem2015}.

This paper takes a different angle of attack to the verification
problem, seeking generality of the type of parametric systems and
focusing on the verification of a particular but essential
property: \emph{deadlock-freedom}. The aim is to come up with
effective methods for checking deadlock-freedom, by overcoming the
complexity blowup stemming from the effective generation of
reachability sets. We briefly describe our approach below.

A system is the composition of a finite number of component instances
of given types, using interactions that follow the
Behaviour-Interaction-Priorities (BIP)
paradigm \cite{BasuBBCJNS11}. To simplify the technical part, we
assume that components and interactions are finite abstractions of
real-life systems. An instance is a finite-state transition system
whose edges are labeled by ports. The instances communicate
synchronously via a number of simultaneous interactions involving a
set of ports each, such that no data is exchanged during
interactions. If the number of instances in the system is fixed and
known in advance, we say that the system is \emph{bounded}, otherwise
it is \emph{parametric}.

\begin{figure}[htbp]
  \centering    
  \vspace*{-\baselineskip}            
  \subfloat[Bounded System\label{fig:ex-intro-static}]{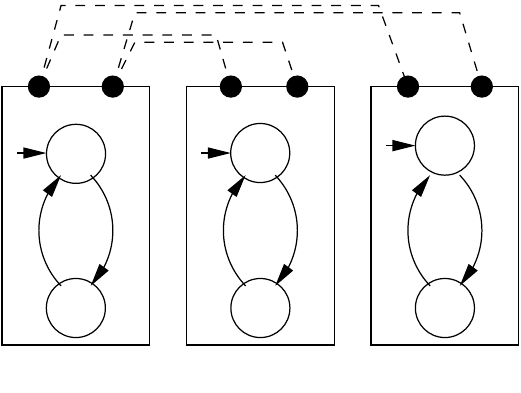}
  \hspace{.3cm}
  \subfloat[Parametric System\label{fig:ex-intro-parametric}]{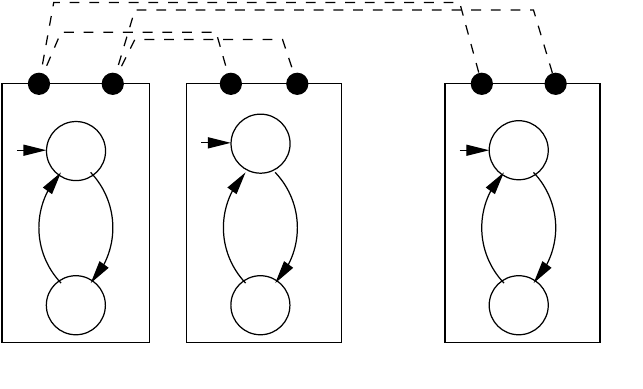}
  \vspace*{-.5\baselineskip}
  \caption{\label{fig:ex-intro} Mutual Exclusion Example}
  \vspace*{-\baselineskip}
\end{figure}

For instance, the bounded system in Figure \ref{fig:ex-intro-static}
consist of component types \emph{Semaphore}, with one instance,
and \emph{Task}, with two instances. A semaphore goes from the free
state $r$ to the taken state $s$ by an acquire action $a$, and
viceversa from $s$ to $r$ by a release action $e$. A task goes from
waiting $w$ to busy $u$ by action $b$ and viceversa, by action
$f$. For the bounded system in Figure \ref{fig:ex-intro-static}, the
interactions are $\set{a,b_1}, \set{a,b_2}, \set{e,f_1}$ and
$\set{e,f_2}$, depicted with dashed lines. Since the number of
instances is known in advance, we can view an interaction as a minimal
satisfying valuation of the boolean formula $\interform = (a\wedge
b_1) \vee (a\wedge b_2) \vee (e\wedge f_1) \vee (e\wedge f_2)$, where
the port symbols are propositional variables. Because every instance
has finitely many states, we can write a boolean formula $\Delta =
[\neg r \vee \neg (w_1 \vee w_2)] \wedge [\neg s \vee \neg (u_1 \vee
u_2)]$, this time over propositional state variables, which defines
the configurations in which all interactions are disabled (deadlock).
Proving that no deadlock configuration is reachable from the initial
configuration $r \wedge w_1 \wedge w_2$, requires finding an
over-approximation (invariant) $I$ of the reachable configurations,
such that the conjunction $I \wedge \Delta$ is not satisfiable.

The basic idea of our method, supported by the \textsc{D-Finder}
deadlock detection tool \cite{BensalemBNS09} for bounded
component-based systems, is to compute an invariant straight from the
interaction formula, without going through costly abstract fixpoint
iterations. The invariants we are looking for are in fact solutions of
a system of boolean constraints $\trapconstraint{\interform}$, of size
linear in the size of $\interform$ (written in DNF). In our example,
$\trapconstraint{\interform} = \bigwedge_{i=1,2} (r \vee
w_i) \leftrightarrow (s \vee u_i)$. Finding the (minimal) solutions of
this constraint can be done, as currently implemented
in \textsc{D-Finder}, by exhaustive model enumeration using a SAT
solver. Here we propose a more efficient solution, which consists in
writing $\trapconstraint{\interform}$ in DNF and remove the negative
literals from each minterm. In our case, this gives the invariant $I =
(r \vee s) \wedge \bigwedge_{i=1,2} (w_i \vee u_i) \wedge (r \vee
u_1 \vee u_2) \wedge (s \vee w_1 \vee w_2)$ and $I \wedge \Delta$ is
proved unsatisfiable using a SAT solver.

The main contribution of this paper is the generalization of this
invariant generation method to the parametric case. To understand the
problem, consider the parametric system from
Figure \ref{fig:ex-intro}, in which a \emph{Semaphore} interacts with
$n$ \emph{Tasks}, where $n>0$ is not known in advance. The
interactions are described by a fragment of first order logic, in
which the ports are either propositional or monadic predicate symbols,
in our case $\interform = a \wedge \exists i ~.~ b(i) \vee
e \wedge \exists i ~.~ f(i)$. This logic, called \emph{Monadic
Interaction Logic} (\mil), is also used to express the constraints
$\trapconstraint{\interform}$ and compute their solutions. In our
case, we obtain $I = (r \vee s) \wedge [\forall i ~.~ w(i) \vee
u(i)] \wedge [r \vee \exists i~.~u(i)] \wedge [s \vee \exists
i~.~w(i)]$. As in the bounded case, we can give a parametric
description of deadlock configurations $\Delta = [\neg
r \vee \neg \exists i ~.~ w(i)] \wedge [\neg s \vee \neg \exists i ~.~
u(i)]$ and prove that $I \wedge \Delta$ is unsatisfiable, using the
decidability of \mil, based on an early small model property result
due to L\"owenheim \cite{Lowenheim15}. In practice, we avoid the model
enumeration suggested by this result and check the satisfiability of
such queries using a decidable theory of sets with cardinality
constraints \cite{KuncakNR06}, available in the CVC4 SMT
solver \cite{BansalRBT16}.

The paper is structured as follows: \S\ref{sec:static} presents
existing results for checking deadlock-freedom of bounded systems
using invariants, \S\ref{sec:parametric} formalizes the approach for
computing invariants using \mil, \S\ref{sec:cardinality} introduces
cardinality constraints for invariant
generation, \S\ref{sec:verification} presents the integration of the
above results within a verification technique for parametric systems
and \S\ref{sec:experiments} reports on preliminary experiments carried
out with a prototype tool. Finally, \S\ref{sec:conclusion} presents
concluding remarks and future work directions. For reasons of space,
all proofs are given in \cite{BIS18Arxiv}. 

\section{Bounded Component-based Systems}
\label{sec:static}

A \emph{component} is a tuple $\acomptype
  = \tuple{\ports, \states, \initstate, \rules}$, where $\ports
  = \set{p,q,r,\ldots}$ is a finite set of \emph{ports}, $\states$ is
  a finite set of
\emph{states}, $\initstate\in\states$ is an initial state and $\rules
\subseteq \states \times \ports \times \states$ is a set of
\emph{transitions} written $s \arrow{p}{} s'$. To simplify
the technical details, we assume there are \emph{no two different
transitions with the same port}, i.e.\ if $s_1 \arrow{p_1}{} s'_1,
s_2 \arrow{p_2}{} s'_2 \in \rules$ and $s_1 \neq s_2$ or $s'_1 \neq
s'_2$ then $p_1 \neq p_2$. In general, this restriction can be lifted,
at the cost of cluttering the presentation.

A \emph{bounded system} $\asys = \tuple{\typeno{\acomptype}{1},
  \ldots, \typeno{\acomptype}{n},\interform}$ consists of a fixed
number ($n$) of components $\typeno{\acomptype}{k} =
\tuple{\typeno{\ports}{k}, \typeno{\states}{k},
  \typeno{\initstate}{k}, \typeno{\rules}{k}}$ and an
\emph{interaction formula} $\interform$, describing the allowed interactions.
Since the number of components is known in advance, we write
interaction formulae using boolean logic over the set of propositional
variables
$\bvars \isdef \bigcup_{k=1}^n(\typeno{\ports}{k} \cup \typeno{\states}{k})$.
Here we intentionally use the names of states and ports as
propositional variables.

A \emph{boolean interaction formula} is either $a \in \bvars$, $f_1
\wedge f_2$ or $\neg f_1$, where $f_i$ are formulae, for
$i=1,2$, respectively. We define the usual shorthands $f_1 \vee f_2
\isdef \neg(\neg f_1 \wedge \neg f_2)$, $f_1 \rightarrow f_2 \isdef
\neg f_1 \vee f_2$, $f_1 \leftrightarrow f_2 \isdef (f_1
\rightarrow f_2) \wedge (f_2 \rightarrow f_1)$. A literal is either a variable 
or its negation and a minterm is a conjunction of literals. A formula
is in disjunctive normal form (DNF) if it is written as
$\bigvee_{i=1}^n\bigwedge_{j=1}^{m_i} \ell_{ij}$, where $\ell_{ij}$ is
a literal. A formula is \emph{positive} if and only if each variable
occurs under an even number of negations, or, equivalently, its DNF
forms contains no negative literals. We assume interaction formulae of
bounded systems to be always positive.

A \emph{boolean valuation} $\beta : \bvars \rightarrow
\set{\true,\false}$ maps each propositional variable to either true ($\true$) 
or false ($\false$). We write $\beta \models f$ if and only if
$f=\true$, when replacing each boolean variable $a$ with $\beta(a)$ in
$f$. We say that $\beta$ is a \emph{model} of $f$ in this case and
write $f \equiv g$ for $\sem{f} = \sem{g}$, where
$\sem{f} \isdef \set{\beta \mid \beta \models f}$.  Given two
valuations $\beta_1$ and $\beta_2$ we write $\beta_1
\subseteq \beta_2$ if and only if $\beta_1(a) = \true$ implies $\beta_2(a) = \true$, 
for each variable $a \in \bvars$. We write $f \minequiv g$ for
$\minsem{f}=\minsem{g}$, where
$\minsem{f}\isdef\set{\beta \in \sem{f} \mid \text{ for all } \beta':~ 
\beta' \subseteq \beta \text{ and } \beta' \neq \beta \text{
only if } \beta' \not\in \sem{f}}$ is the set of minimal models of
$f$.

\subsection{Execution Semantics of Bounded Systems} 

We use 1-safe marked Petri Nets to define the set of executions of a
bounded system. A \emph{Petri Net} (PN) is a tuple $\anet =
\tuple{\places,\trans,\edges}$, where $\places$ is a set of
\emph{places}, $\trans$ is a set of \emph{transitions}, $\places \cap
\trans = \emptyset$, and $\edges \subseteq \places \times \trans \cup
\trans \times \places$ is a set of \emph{edges}. The elements of
$\places \cup \trans$ are called \emph{nodes}. For a node $n$, let
  $\pre{n} \isdef \set{m \in \places \cup \trans \mid E(m,n)=1}$,
  $\post{n} \isdef \set{m \in \places \cup \trans \mid E(n,m)=1}$ and
  lift these definitions to sets of nodes, as usual.

\begin{wrapfigure}{r}{.4\textwidth}
  \centering    
  \vspace*{-2\baselineskip}            
  \scalebox{.8}{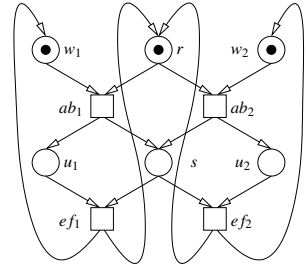}
  \caption{PN for Mutual Exclusion}
  \label{fig:pn-example} 
  \vspace*{-1.5\baselineskip}
\end{wrapfigure}

A \emph{marking} for a PN $\anet = \tuple{\places,\trans,\edges}$ is a
function $\amark : \places \rightarrow \nat$.
A \emph{marked Petri net} is a pair $\amarkednet=(\anet,\amark_0)$,
where $\amark_0$ is the \emph{initial marking} of $\anet =
\tuple{\places,\trans,\edges}$. We consider that the
reader is familiar with the standard execution semantics of a marked
PN. A marking $\amark$ is \emph{reachable} in $\amarkednet$ if and
only if there exists a sequence of transitions leading fom $\amark_0$
to $\amark$. We denote by $\reach{\amarkednet}$ the set of reachable
markings of $\amarkednet$. A set of markings $\mathcal{M}$ is
an \emph{invariant} of $\amarkednet=(\anet,\amark_0)$ if and only if
$\amark_0 \in \mathcal{M}$ and $\mathcal{M}$ is closed under the
transitions of $\anet$. A marked PN $\amarkednet$ is \emph{$1$-safe}
if $\amark(s) \leq 1$, for each $s \in \places$ and each
$\amark \in \reach{\amarkednet}$. In the following, we consider only
marked PNs that are $1$-safe. In this case, any (necessarily finite)
set of reachable markings can be defined by a boolean formula, which
identifies markings with the induced boolean valuations.  A marking
$\amark$ is a \emph{deadlock} if for no transition is enaled in
$\amark$ and let $\dead{\amarkednet}$ be the set of deadlocks of
$\anet$. A marked PN $\amarkednet$ is \emph{deadlock-free} if and only
if $\reach{\amarkednet} \cap \dead{\amarkednet} = \emptyset$. A
sufficient condition for deadlock freedom is $\mathcal{M} \cap
\dead{\amarkednet} = \emptyset$, for some invariant $\mathcal{M}$ of
$\amarkednet$. 

In the rest of this section, we fix a bounded system $\asys
= \tuple{\typeno{\acomptype}{1}, \ldots, \typeno{\acomptype}{n},\interform}$,
where $\typeno{\acomptype}{k} =
\tuple{\typeno{\ports}{k}, \typeno{\states}{k},
  \typeno{\initstate}{k}, \typeno{\rules}{k}}$, for all $k \in [1,n]$
  and $\interform$ is a positive boolean formula, over propositional
  variables denoting ports. The set of executions of $\asys$ is given
  by the 1-safe marked PN $\amarkednet_\asys = (\anet,\amark_0)$,
  where $\anet=(\bigcup_{i=1}^n \typeno{\states}{i},\trans,\edges)$,
  $\amark_0(s)=1$ if and only if
  $s \in \set{\typeno{\initstate}{i} \mid i\in[1,n]}$ and $\trans$,
  $\edges$ are as follows. For each minimal model
  $\beta \in \minsem{\interform}$, we have a transition
  $\atrans_\beta \in T$ and edges $(s_i, \atrans_\beta),
  (\atrans_\beta, s'_i) \in E$, for all $i \in [1,n]$ such that
  $s_i \arrow{p_i}{} s'_i \in \typeno{\rules}{i}$ and $\beta(p_i)
  = \true$. Moreover, nothing else is in $T$ or $E$.

For example, the marked PN from Figure \ref{fig:pn-example} describes
the set of executions of the bounded system from
Figure \ref{fig:ex-intro-static}. Note that each transition of the PN
corresponds to a minimal model of the interaction formula $\interform
= a\wedge b_1 \vee a\wedge b_2 \vee e\wedge f_1 \vee e\wedge f_2$, or
equivalently, to the set of (necessarily positive) literals of some
minterm in the DNF of $\interform$.

\subsection{Proving Deadlock Freedom of Bounded Systems}

A bounded system $\asys$ is deadlock-free if and only if its
corresponding marked PN $\amarkednet_\asys$ is deadlock-free. In the
following, we prove deadlock-freedom of a bounded system, by defining
a class of invariants that are particularly useful for excluding
unreachable deadlock markings.

Given a Petri Net $\anet = (\places, \trans, \edges)$, a set of places
$W \subseteq \places$ is called a \emph{trap} if and only if
$\post{W} \subseteq \pre{W}$. A trap $W$ of $\anet$ is a \emph{marked
trap} of the marked PN $\amarkednet = (\anet,\amark_0)$ if and only if
$\amark_0(s)=\true$ for some $s \in W$. A \emph{minimal marked trap}
is a marked trap such that none of its strict subsets is a marked
trap. A marked trap defines an invariant of the PN because some place
in the trap will always be marked, no matter which transition is
fired. The \emph{trap invariant} of $\amarkednet$ is the least set of
markings that mark each trap of $\amarkednet$. Clearly, the trap
invariant of $\amarkednet$ subsumes the set of reachable markings of
$\amarkednet$, because the latter is the least invariant of
$\amarkednet$ and invariants are closed under
intersection\footnote{The intersection of two or more invariants is
again an invariant.}.

\begin{lemma}\label{lemma:trap-invariant}
  Given a bounded system $\asys$, the boolean
  formula: \[\begin{array}{c}
  \alltrap{\amarkednet_\asys} \isdef \bigwedge\set{\bigvee_{i=1}^k
  s_i \mid \set{s_1,\ldots,s_k} \text{ is a marked trap of
  } \amarkednet_\asys}
  \end{array}\] defines an invariant of $\amarkednet_\asys$.
\end{lemma}
\proof{ Let $\amarkednet_\asys = (\anet, \amark_0)$, where $\anet =
  (\places, \trans, \edges)$. First, we prove that $\amark_0 \models
  \alltrap{\amarkednet_\asys}$. Let $S = \set{s_1, \ldots, s_k}$ be a
  marked trap of $\amarkednet_\asys$. Since $S$ is marked,
  $\amark_0(s_i) = \true$ for some $i \in [1,k]$, thus $\amark_0
  \models \bigvee_{i=1}^k s_i$. Because the choice of $S$ is
  arbitrary, we have $\amark_0 \models
  \alltrap{\amarkednet_\asys}$. Second, let $\amark \in
  \sem{\alltrap{\amarkednet_\asys}}$ and $t \in T$ such that $\amark
  \arrow{t}{} \amark'$. We prove that $\amark' \models
  \alltrap{\amarkednet_\asys}$. Let $S = \set{s_1, \ldots, s_k}$ be a
  marked trap of $\amarkednet_\asys$. Then $\amark \models s_i$ for
  some $i \in [1,k]$ and, because $S$ is a trap, $\amark' \models s_j$
  for some $j \in [1,k]$. Since the choice of $S$ was arbitrary, we
  obtain $\amark' \models \alltrap{\amarkednet_\asys}$. \qed}

Next, we describe a method of computing trap invariants that does not
explicitly enumerate all the marked traps of a marked PN. First, we
consider a \emph{trap constraint} $\trapconstraint{\interform}$,
derived from the interaction formula $\interform$, in linear time. By
slight abuse of notation, we define, for a given port $p \in
\typeno{\ports}{i}$ of the component $\typeno{\acomptype}{i}$, for
some $i \in [1,n]$, the pre- and post-state of $p$ in
$\typeno{\acomptype}{i}$ as $\pre{p} \isdef s$ and $\post{p} \isdef
s'$, where $s \arrow{p}{} s'$ is the unique rule\footnote{We have
  assumed that each port is associated a unique transition rule.}
involving $p$ in $\typeno{\rules}{i}$, and $\pre{p} = \post{p} \isdef
\false$ if there is no such rule. Assuming that the interaction formula 
is written in DNF as $\interform
= \bigvee_{k=1}^N\bigwedge_{\ell=1}^{M_k} p_{k\ell}$, we define the
trap constraint:
\[\begin{array}{c}
\trapconstraint{\interform} \isdef \bigwedge_{k=1}^N
\left(\bigvee_{\ell=1}^{M_k} \pre{p_{k\ell}}\right) \rightarrow
\left(\bigvee_{\ell=1}^{M_k} \post{p_{k\ell}}\right)
\end{array}\]
It is not hard to show\footnote{See \cite{BarkaouiLemaire89} for a
proof.} that any satisfying valuation of $\trapconstraint{\interform}$
defines a trap of $\amarkednet_\asys$ and, moreover, any such trap is
defined in this way. We also consider the formula
$\init{\asys} \isdef \bigvee_{k=1}^n \typeno{s_0}{k}$ defining the set
of initially marked places of $\asys$, and prove the following:

\begin{lemma}\label{lemma:static-trap-constraint}
Let $\asys$ be a bounded system with interaction formula $\interform$
and $\beta$ be a boolean valuation. Then
$\beta \in \sem{\trapconstraint{\interform} \wedge \init{\asys}}$ iff
$\set{s \mid \beta(s) = \true}$ is a marked trap of
$\amarkednet_\asys$.  Moreover,
$\beta \in \minsem{\trapconstraint{\interform} \wedge \init{\asys}}$
iff $\set{s \mid \beta(s) = \true}$ is a minimal marked trap of
$\amarkednet_\asys$.
\end{lemma}
\proof{ Let $\typeno{\acomptype}{i} = \tuple{\typeno{\ports}{i},
    \typeno{\states}{i}, \typeno{\initstate}{i}, \typeno{\rules}{i}}$,
  for all $i \in [1,n]$, $\amarkednet_\asys = (\anet,\amark_0)$ and
  $\anet = \tuple{\places, \trans, \edges}$, where $Q =
  \bigcup_{i=1}^n \typeno{\states}{i}$ and $T = \set{\atrans_\beta
    \mid \beta \in \minsem{\interform}}$. Given a trap $S \subseteq Q$
  of $\anet$, we have the following equivalences:
  \[\begin{array}{rcll}
  \post{S} & \subseteq & \pre{S} & \iff \\
  \bigwedge_{s \in \places} [s \in S & \rightarrow & \set{t \in \trans \mid (s,t) \in \edges} \subseteq \set{t \in \trans \mid (t,s) \in \edges}] & \iff \\
  \bigwedge_{s \in \places} [s \in S & \rightarrow & (\bigwedge_{\atrans \in \trans} s \in \pre{\atrans} \rightarrow \bigvee_{s' \in \places} s' \in S \wedge s' \in \post{\atrans})] & \iff \\
  \bigwedge_{s \in \places} \bigwedge_{\atrans \in \trans} (s \in S \wedge s \in \pre{\atrans} & \rightarrow & \bigvee_{s' \in \places} s' \in S \wedge s' \in \post{\atrans}) & \iff \\
  \bigwedge_{\atrans \in \trans} (\bigvee_{\text{$s \in \pre{\atrans}$}} s \in S & \rightarrow & \bigvee_{s' \in \post{\atrans}} s' \in S)
  \end{array}\]
  Assume that $[s \in \places]$ is a propositional variable. Then for each
  transition $\atrans_\beta \in T$, we have:
  \[\begin{array}{rcl}
  \bigvee_{\text{$s \in \pre{\atrans_\beta}$}} [s\in S] & \iff & \bigvee_{\beta(p) = \true} \pre{p} \\
  \bigvee_{\text{$s \in \post{\atrans_\beta}$}} [s\in S] & \iff & \bigvee_{\beta(p) = \true} \post{p}
  \end{array}\]
  Clearly, for any valuation $\beta \in
  \sem{\trapconstraint{\interform}}$ of the propositional variables
  corresponding to the places in $Q$ that satisfies
  $\trapconstraint{\interform}$, the set $S_\beta = \set{s \mid
    \beta(s) = \true}$ is a trap of $N$. If, moreover, $\beta \models
  \bigvee_{i=1}^n \typeno{\initstate}{i}$ then $S_\beta$ is a marked
  trap of $\amarkednet_\asys$. Furthermore, $\beta$ is a minimal model
  of $\trapconstraint{\interform} \wedge \bigwedge_{i=1}^n
  \typeno{s_0}{i}$ iff for each valuation $\beta' \subseteq \beta$, 
  such that $\beta'\neq\beta$, we have $\beta' \not\models \trapconstraint{\interform} 
  \wedge \bigvee_{i=1}^n \typeno{s_0}{i}$. But then, no strict subset of
  $\set{s \mid \beta(s) = \true}$ is a marked trap of
  $\amarkednet_\asys$, thus $\set{s \mid \beta(s) = \true}$ is a
  minimal marked trap of $\amarkednet_\asys$. \qed}

Because $\trapconstraint{\interform}$ and $\init{\asys}$ are boolean
formulae, it is, in principle, possible to compute the trap invariant
$\alltrap{\amarkednet_\asys}$ by enumerating the (minimal) models of
$\trapconstraint{\interform} \wedge \init{\asys}$ and applying the
definition from Lemma \ref{lemma:trap-invariant}. However, model
enumeration is inefficient and, moreover, does not admit
generalization for the parametric case, in which the size of the
system is unknown. For these reasons, we prefer a computation of the
trap invariant, based on two symbolic transformations of boolean
formulae, described next.

For a formula $f$ we denote by $\pos{f}$ the positive formula obtained
by deleting all negative literals from the DNF of $f$. We shall call
this operation \emph{positivation}. Second, for a positive boolean
formula $f$, we define the \emph{dual} formula $\dual{f}$ recursively
on the structure of $f$, as follows: $\dual{f_1 \wedge
f_2} \isdef \dualnopar{f_1} \vee \dualnopar{f_2}$, $\dual{f_1 \vee
f_2} \isdef \dualnopar{f_1} \wedge \dualnopar{f_2}$ and
$\dualnopar{a} \isdef a$, for any $a \in \bvars$. Note that
$\dualnopar{f}$ is equivalent to the negation of the formula obtained
from $f$ by substituting each variable $a$ with $\neg a$ in $f$.

\ifLongVersion
\begin{lemma}\label{lemma:boolean-pos-dual}
  Given boolean formulae $f$ and $g$, we have $f \equiv g$ only if
  $\dual{\pos{f}} \equiv \dual{\pos{g}}$.
\end{lemma}
\proof{If $f \equiv g$, the set of minterms in the DNF of $f$
is identical to the one of $g$, modulo commutativity of
conjunctions. Then the set of minterms in the DNF of $\pos{f}$ equals
the one of $\pos{g}$, thus $\pos{f} \equiv \pos{g}$.  Second, the CNF
of $\dual{\pos{f}}$ is the same of the CNF of $\dual{\pos{g}}$, as
both are obtained directly from the DNF of $\pos{f}$ and $\pos{g}$,
respectively, by interchanging disjunctions with conjunctions. \qed}
\fi

The following theorem gives the main result of this section, the
symbolic computation of the trap invariant of a bounded system,
directly from its interaction formula.

\begin{theorem}\label{thm:static-trap-inv}
  For any bounded system $\asys$, with interaction formula
  $\interform$, we have: 
  \[\begin{array}{c}
  \alltrap{\amarkednet_\asys} \equiv \dual{\pos{\left[\trapconstraint{\interform} \wedge \init{\asys}\right]}}
  \end{array}\]
\end{theorem}
\proof{ For a boolean valuation $\beta$, we denote by $\mu_\beta$ the
  complete minterm $\bigwedge_{\beta(a)=\true} a \wedge
  \bigwedge_{\beta(a)=\false} \neg a$. By Lemma
  \ref{lemma:static-trap-constraint} we obtain the equivalence: 
  \[\trapconstraint{\interform} \wedge
  \init{\amarkednet_\asys} \equiv \bigvee \Big\{\mu_\gamma \mid \set{s \mid \gamma(s)=\true} \text{ is a marked trap of } \amarkednet\Big\}\]
  and thus:
  \[\begin{array}{rcl}
  \pos{[\trapconstraint{\interform} \wedge \init{\amarkednet_\asys}]} & \equiv & 
    \bigvee \Big\{\pos{\mu_\gamma} \mid \set{s \mid \gamma(s)=\true} \text{ is a marked trap of } \amarkednet_\asys\Big\} \\
    & \equiv & \bigvee \Big\{ \bigvee_{i=1}^k s_i \mid \set{s_1,\ldots,s_k} \text{ is a marked trap of } \amarkednet_\asys\Big\} \\
    & \equiv & \dual{\alltrap{\amarkednet_\asys}}
    \end{array}\]
  The equivalence of the statement is obtained by applying Lemma
  \ref{lemma:boolean-pos-dual}. \qed}
  
Intuitively, any satisfying valuation
of \(\trapconstraint{\interform} \wedge \init{\asys}\) defines an
initially marked trap of $\amarkednet_\asys$ and a minimal such
valuation defines a minimal such trap
(Lemma \ref{lemma:static-trap-constraint}). Instead of computing the
minimal satisfying valuations by model enumeration, we directly cast
the above formula in DNF and remove the negative literals. This is
essentially because the negative literals do not occur in the
propositional definition of a set of places\footnote{If the DNF is
$(p \wedge q) \vee (p \wedge \neg r)$, the dualization would give
$(p \vee q) \wedge (p \vee \neg r)$. The first clause corresponds to
the trap $\set{p,q}$ (either $p$ or $q$ is marked), but the second
does not directly define a trap. However, by first removing the
negative literals, we obtain the traps $\set{p,q}$ and
$\set{r}$.}. Then the dualization of this positive formula yields the
trap invariants in CNF, as a conjunction over disjunctions of
propositional variables corresponding to the places inside a minimal
initially marked trap.

Just as any invariants, trap invariants can be used to prove absence
of deadlocks in a bounded system. Assuming, as before, that the
interaction formula is given in DNF as $\interform =
\bigvee_{k=1}^N\bigwedge_{\ell=1}^{M_k} p_{k\ell}$, we define the set of 
deadlock markings of $\amarkednet_\asys$ by the formula 
\(\deadconstraint{\interform} \isdef \bigwedge_{k=1}^N \bigvee_{\ell=1}^{M_k}
  \neg(\pre{p_{k\ell}})\). This is the set of configurations in which
  all interactions are disabled. With this definition, proving
  deadlock freedom amounts to proving unsatisfiability of a boolean
  formula.

\begin{corollary}\label{cor:static-deadlock-freedom}
  A bounded  system $\asys$ with interaction formula
  $\interform$ is deadlock-free if the boolean
  formula \(\dual{\pos{[\trapconstraint{\interform} \wedge \init{\asys}]}} \wedge \deadconstraint{\interform}\)
  is unsatisfiable.
\end{corollary}
\proof{
Let $\amarkednet_\asys = (\anet,\amark_0)$, where
$\anet=(\places,\trans,\edges)$ and define the set of deadlock
markings:
\[\alldead{\amarkednet_\asys} \isdef \bigwedge_{t \in T} \bigvee_{\text{$s \in \pre{t}$}} \neg s\]
Suppose, by contradiction, that $\asys$ is not deadlock-free, thus
$\reach{\amarkednet_\asys} \wedge \alldead{\amarkednet_\asys}$ has a
satisfying valuation $\beta$. Because $\alltrap{\amarkednet_\asys}$
defines an invariant of $\amarkednet_\asys$ and
$\reach{\amarkednet_\asys}$ defines its least invariant, we have
$\reach{\amarkednet_\asys} \rightarrow \alltrap{\amarkednet_\asys}$
thus $\beta \models \alltrap{\amarkednet_\asys} \wedge
\alldead{\amarkednet_\asys}$.  By Theorem \ref{thm:static-trap-inv},
we have \(\alltrap{\amarkednet_\asys} \rightarrow
\dual{\pos{[\trapconstraint{\interform} \wedge
      \init{\amarkednet_\asys}]}}\) and, from the definition of
$\amarkednet_\asys$, one also obtains that
\(\alldead{\amarkednet_\asys} \rightarrow
\deadconstraint{\interform}\) leading to $\beta \models
\dual{\pos{[\trapconstraint{\interform} \wedge
      \init{\amarkednet_\asys}]}} \wedge
\deadconstraint{\interform}$, which contradicts
\(\dual{\pos{[\trapconstraint{\interform} \wedge
      \init{\amarkednet_\asys}]}} \wedge \deadconstraint{\interform}
\rightarrow \false\). \qed
}

\section{Parametric Component-based Systems}
\label{sec:parametric}

From now on we shall focus on parametric systems,
consisting of a fixed set of component types
$\typeno{\acomptype}{1}, \ldots, \typeno{\acomptype}{n}$, such that
the number of instances of each type is not known in advance. These
numbers are given by a function $\maxnf : [1,n] \rightarrow \nat$,
where $\maxnf(k)$ denotes the number of components of type
$\typeno{\acomptype}{k}$ that are active in the system. To simplify
the technical presentation of the results, we assume that all
instances of a component type are created at once, before the system
is started\footnote{This is not a limitation, since dynamic instance
creation can be simulated by considering that all instances are
initially in a waiting state, which is left as result of an
interaction involving a designated ``spawn'' port.}. For the rest of
this section, we fix a parametric system $\asys
= \tuple{\typeno{\acomptype}{1}, \ldots, \typeno{\acomptype}{n}, \maxnf, \interform}$,
where each component type $\typeno{\acomptype}{k}
= \tuple{\typeno{\ports}{k}, \typeno{\states}{k}, \typeno{\initstate}{k}, \typeno{\rules}{k}}$
has the same definition as a component in a bounded system and
$\interform$ is an interaction formula, written in the fragment of
first order logic, defined next.

\subsection{Monadic Interaction Logic}
\label{sec:mil}

For each component type $\typeno{\acomptype}{k}$, where $k \in [1,n]$,
we assume a set of index variables $\typeno{\vars}{k}$ and a set of
predicate symbols $\typeno{\preds}{k} \isdef \typeno{\ports}{k} \cup
\typeno{\states}{k}$. Similar to the bounded case, we use state and
ports names as monadic (unary) predicate symbols. We also define the
sets $\vars \isdef \bigcup_{k=1}^n \typeno{\vars}{k}$ and
$\preds \isdef \bigcup_{k=1}^n \typeno{\preds}{k}$. Moreover, we
consider that $\typeno{\vars}{k} \cap \typeno{\vars}{\ell}
= \emptyset$ and $\typeno{\preds}{k} \cap \typeno{\preds}{\ell}
= \emptyset$, for all $ 1 \leq k < \ell \leq n$. For simplicity's
sake, we assume that all predicate symbols in $\preds$ are of arity
one. For component types $\typeno{\acomptype}{k}$, such that
$\maxnf(k)=1$ and predicate symbols $\apred \in \typeno{\preds}{k}$,
we shall write $\apred$ instead of $\apred(1)$, as in the interaction
formula of the system from Figure \ref{fig:ex-intro-parametric}. The
syntax of the \emph{monadic interaction logic} (\mil) is given below:
\[\begin{array}{rcl}
i,j & \in & \vars ~~\text{ index variables} \\
\phi & := & i=j \mid \apred(i) \mid \phi_1 \wedge \phi_2
\mid \neg \phi_1 \mid \exists i ~.~ \phi_1 
\end{array}\]
where, for each predicate atom $\apred(i)$, if
$\apred \in \typeno{\preds}{k}$ and $i \in \typeno{\vars}{\ell}$ then
$k=\ell$. We use the shorthands $\forall i
~.~ \phi_1 \isdef \neg(\exists i ~.~ \neg\phi_1)$
and \(\distinct(i_1,\ldots,i_m) \isdef \bigwedge_{1 \leq j < \ell \leq m}
 \neg i_j = i_\ell\)\footnote{Throughout this paper, we consider
that $\bigwedge_{i \in I} \phi_i = \true$ if
$I=\emptyset$.}. A \emph{sentence} is a formula in which all variables
are in the scope of a quantifier. A formula is
\emph{positive} if each predicate symbol occurs under an even number of
negations. The semantics of \mil\ is given in terms of structures $\I
= (\U,\nu,\iota)$, where: 
\begin{compactitem}
\item $\U \isdef [1,\max_{k=1}^n \maxn{k}]$ is the \emph{universe}
  of instances, over which variables range,
\item $\nu : \vars \rightarrow \U$ is a
  \emph{valuation} mapping variables to elements of the
  universe,
\item $\iota : \preds \rightarrow 2^{\U}$
  is an \emph{interpretation} of predicates as subsets of the
  universe.
\end{compactitem}
For a structure $\I = (\U,\nu,\iota)$ and a formula $\phi$, the
satisfaction relation $\I \models \phi$ is defined as:
 \[\begin{array}{lclclcll}
 \I \models \bot & \Leftrightarrow & \text{never} & \hspace*{2mm} & \I \models i=j & \Leftrightarrow & \nu(i)=\nu(j) \\
 \I \models p(i) & \Leftrightarrow & \nu(i) \in \iota(p) &&  
 \I \models \exists i ~.~ \phi_1 & \Leftrightarrow & (\U,\nu[i \leftarrow m],\iota) \models \phi_1 
 & \text{  for some $m \in [1,\maxn{k}]$} \\
 &&&&&&& \text{  provided that $i \in \typeno{\vars}{k}$}
 \end{array}\]
where $\nu[i \leftarrow m]$ is the valuation that acts as $\nu$,
except for $i$, which is assigned to $m$. Whenever $\I \models \phi$,
we say that $\I$ is a \emph{model} of $\phi$. It is known that, if a 
\mil\ formula has a model, then it has a model with universe of cardinality 
at most exponential in the size (number of symbols) of the
formula \cite{Lowenheim15}. This result, due to L\"owenheim, is among
the first decidability results for a fragment of first order logic.

Structures are partially ordered by pointwise inclusion, i.e.\ for
$\I_i=(\U,\nu_i,\iota_i)$, for $i=1,2$, we write $\I_1 \subseteq \I_2$
iff $\iota_1(p) \subseteq \iota_2(p)$, for all $p \in \preds$ and
$\I_1 \subset \I_2$ iff $\I_1 \subseteq \I_2$ and $\I_1 \neq \I_2$. As
before, we define the sets $\sem{\phi}
= \set{\I \mid \I \models \phi}$ and $\minsem{\phi}
= \set{\I \in \sem{\phi} \mid \forall \I'
~.~ \I' \subset \I \rightarrow \I' \not\in \sem{\phi}}$ of models and
minimal models of a \mil\ formula, respectively. Given formulae
$\phi_1$ and $\phi_2$, we write $\phi_1
\equiv \phi_2$ for $\sem{\phi_1} = \sem{\phi_2}$ and $\phi_1 \minequiv
\phi_2$ for $\minsem{\phi_1} = \minsem{\phi_2}$.

\subsection{Execution Semantics of Parametric Systems}

We consider the interaction formulae of parametric systems to be
finite disjunctions of formulae of the form below:
\begin{equation}\label{eq:param-interform}
\begin{array}{c}
\exists i_1 \ldots \exists i_\ell \wedge \varphi \wedge \bigwedge_{j=1}^\ell 
p_j(i_j) \wedge \bigwedge_{j=\ell+1}^{\ell+m} \forall i_j ~.~ \psi_j \rightarrow p_j(i_j)
\end{array}
\end{equation}
where $\varphi,\psi_{\ell+1}, \ldots, \psi_{\ell+m}$ are conjunctions
of equalities and disequalities involving index
variables. Intuitively, the formulae (\ref{eq:param-interform}) state
that there are at most $\ell$ component instances that engage in a
multiparty rendez-vous interaction on ports $p_1(i_1), \ldots,
p_\ell(i_\ell)$, together with a broadcast to the ports
$p_{\ell+1}(i_{\ell+1}), \ldots, p_{\ell+m}(i_{\ell+m})$ of the
instances that fulfill the constraints
$\psi_{\ell+1}, \ldots, \psi_{\ell+m}$. Observe that, if $m=0$, the
above formula corresponds to a multiparty (generalized) rendez-vous
interaction \(\exists i_1 \ldots \exists
i_\ell \wedge \varphi \wedge \bigwedge_{j=1}^\ell p_j(i_j)\).  An
example of peer-to-peer rendez-vous is the parametric system from
Figure \ref{fig:ex-intro}. Another example of broadcast is given
below.

\begin{example}\label{ex:bc}
Consider the parametric system obtained from an arbitrary number of
\emph{Worker} components (Figure \ref{fig:bip-broadcast}), where 
$\typeno{\acomptype}{1} = \mathit{Worker}$, $\typeno{\vars}{1}
= \set{i,i_1,i_2,j}$ and $\typeno{\preds}{1} = \set{a,b,f,u,w}$. Any
pair of instances can jointly execute the $b$ ({\it begin}) action
provided {\em all} others are taking the $a$ ({\it await}) action.
Any instance can also execute alone the $f$ ({\it finish}) action.

\begin{figure}[htbp]
\vspace*{-\baselineskip}
\begin{center}
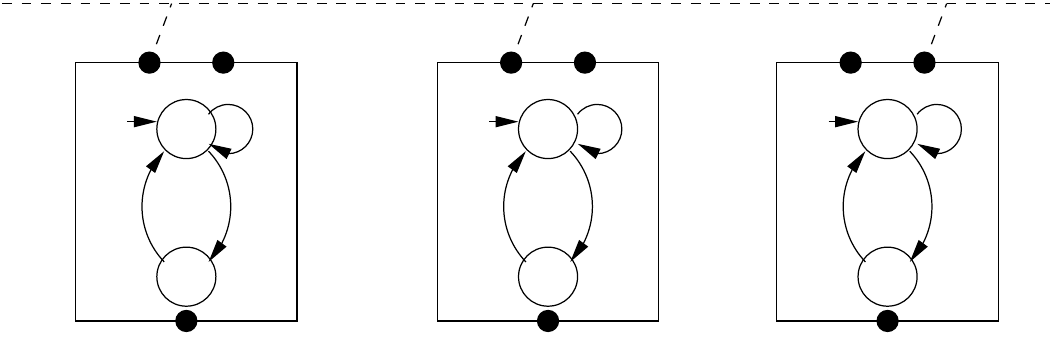
\vspace*{-.5\baselineskip}
\[\interform = [\exists i_1 \exists i_2 ~.~ i_1\not=i_2 \wedge b(i_1) \wedge b(i_2) \wedge \forall j ~.~ 
j\not=i_1 \wedge j\not=i_2 \rightarrow  a(j)] ~\vee~ \exists i. f(i)\]
\end{center}
\vspace*{-\baselineskip}
\caption{Parametric System with Broadcast\label{fig:bip-broadcast}}
\vspace*{-\baselineskip}
\end{figure}
\end{example}

The execution semantics of a parametric system $\asys$ is the marked
PN $\amarkednet_\asys = (\anet,\amark_0)$, where $N =
(\bigcup_{k=1}^n \typeno{\states}{k}
\times [1,\maxn{k}], T, E)$, $\amark_0((\typeno{\initstate}{k}, i)) = 1$, 
for all $k \in [1,n]$ and $i \in [1,\maxn{k}]$, and the sets of
transitions $T$ and edges $E$ are defined next. For each minimal model
$\I = (\U,\nu,\iota) \in \minsem{\interform}$, we have a transition
$\atrans_\I \in T$ and the edges $((s_i,k),\atrans_\I),
(\atrans_\I,(s'_i,k)) \in E$ for all $i \in [1,n]$ such that
$s_i \arrow{p_i}{} s'_i \in \typeno{\rules}{i}$ and
$k \in \iota(p_i)$. Moreover, nothing else is in $T$ or $E$.

As a remark, unlike in the case of bounded systems, the size of the
marked PN $\amarkednet_\asys$, that describes the execution semantics
of a parametric system $\asys$, depends on the maximum number of
instances of each component type. The definition of the trap invariant
$\alltrap{\amarkednet_\asys}$ is the same as in the bounded case,
except that, in this case, the size of the boolean formula depends on
the (unbounded) number of instances in the system. The challenge,
addressed in the following, is to define trap invariants using \mil\
formulae of a fixed size. 

\subsection{Computing Parametric Trap Invariants}

To start with, we define the trap constraint of an interaction formula
 $\interform$ consisting of a finite disjunction of
 (\ref{eq:param-interform}) formulae, as a finite conjunction of
 formulae of the form below:
\[\begin{array}{rcl}
\forall i_1 \ldots \forall i_\ell & . & \left[\varphi \wedge 
\left(\bigvee_{j=1}^\ell \pre{p_j}(i_j) \vee \bigvee_{j=\ell+1}^{\ell+m} \exists i_j ~.~ \psi_j \wedge \pre{p_j}(i_j)\right)\right]
\rightarrow \\[2mm]
&& \left[\bigvee_{j=1}^\ell \post{p_j}(i_j) \vee \bigvee_{j=\ell+1}^{\ell+m} \exists i_j ~.~ \psi_j \wedge \post{p_j}(i_j)\right]
\end{array}\]
where, for a port $p \in \typeno{\ports}{k}$ of some component type
$\typeno{\acomptype}{k}$, $\pre{p(i)}$ and $\post{p(i)}$ denote the
unique predicate atoms $s(i)$ and $s'(i)$, such that $s \arrow{p}{}
s' \in \typeno{\rules}{k}$ is the (unique) transition involving $p$ in
$\typeno{\trans}{k}$, or $\bot$ if there is no such rule.

\begin{example}\label{ex:tc}
For example, the trap constraint for the parametric (rendez-vous) system in Figure \ref{fig:ex-intro-parametric} is
\(\forall i. [r \vee w(i)] \rightarrow [s \vee u(i)] ~\wedge~ \forall i. [s \vee u(i)]\rightarrow [r \vee u(i)]\). 
Analogously, the trap constraint for the parametric (broadcast) system in Figure \ref{fig:bip-broadcast} is: 
\[\begin{array}{rl}
\forall i_1. \forall i_2. & 
[i_1 \not= i_2 \wedge \left(w(i_1) \vee w(i_2) \vee \exists j. (j\not= i_1 \wedge j\not= i_2 \wedge w(j))\right)] \rightarrow \\
& [i_1 \not= i_2 \wedge \left(u(i_1) \vee u(i_2) \vee \exists j. (j\not= i_1 \wedge j\not= i_2 \wedge w(j))\right)]  \\
\wedge ~~ \forall i. & u(i) \rightarrow w(i)
\end{array} \]
\end{example}

We define a translation of \mil\ formulae into boolean formulae of
unbounded size. Given a function $\maxnf : [1,n] \rightarrow \nat$,
the
\emph{unfolding} of a \mil\ sentence $\phi$ is the boolean formula
$\bool{\maxnf}{\phi}$ obtained by replacing each existential
[universal] quantifier $\exists i ~.~ \psi(i)$ [$\forall i
~.~ \psi(i)$], for $i \in \typeno{\vars}{k}$, by a finite disjunction
[conjunction] $\bigvee_{\ell=1}^{\maxnf(k)} \psi[\ell/i]$
[$\bigwedge_{\ell=1}^{\maxnf(k)} \psi[\ell/i]$], where the
substitution of the constant $\ell \in \maxnf(k)$ for the variable $i$
is defined recursively as usual, except for $\apred(i)[\ell/i] \isdef
(\apred,\ell)$, which is a propositional variable. Further, we relate
structures to boolean valuations of unbounded sizes. For a structure
$\I = (\U,\nu,\iota)$ we define the boolean valuation
$\beta_\I((\apred,\ell)) = \true$ if and only if
$\ell \in \iota(\apred)$, for each predicate symbol $\apred$ and each
integer constant $\ell$. Conversely, for each valuation $\beta$ of the
propositional variables $(\apred,\ell)$, there exists a structure
$\I_\beta = (\U,\nu,\iota)$ such that
$\iota(\apred) \isdef \set{\ell \mid \beta((\apred,\ell))=\true}$, for
each $\apred \in
\preds$. The following lemma relates the semantics of \mil\ formulae with 
that of their boolean unfoldings:

\begin{lemma}\label{lemma:predicate-boolean}
  Given a \mil\ sentence $\phi$ and a function $\maxnf :
  [1,n] \rightarrow \nat$, the following
  hold: \begin{compactenum} \item\label{it1:predicate-boolean} for
  each structure $\I \in \sem{\phi}$, we have
  $\beta_\I \in \sem{\bool{\maxnf}{\phi}}$ and conversely, for each
  valuation $\beta \in \sem{\bool{\maxnf}{\phi}}$, we have
  $\I_\beta \in \sem{\phi}$.
  \item\label{it2:predicate-boolean} for each structure $\I \in
    \minsem{\phi}$, we have $\beta_\I \in
    \minsem{\bool{\maxnf}{\phi}}$ and conversely, for each valuation
    $\beta \in \minsem{\bool{\maxnf}{\phi}}$, we have $\I_\beta \in
    \minsem{\phi}$.
  \end{compactenum}
\end{lemma}
\proof{
(\ref{it1:predicate-boolean}) By induction on the
structure of $\phi$. (\ref{it2:predicate-boolean}) First, it is
routine to prove that, for any two structures $\I_1 \subseteq \I_2$,
we have $\beta_{\I_1} \subseteq \beta_{\I_2}$ and, conversely, for any
two valuations $\beta_1 \subseteq \beta_2$, we have $\I_{\beta_1}
\subseteq \I_{\beta_2}$. Next, let $\I \in \minsem{\phi}$. By the
point (\ref{it1:predicate-boolean}), we have $\beta_\I \in
\sem{\bool{\maxnf}{\phi}}$. Suppose $\beta_\I \not\in
\minsem{\bool{\maxnf}{\phi}}$, which means that there exists $\beta'
\subsetneq \beta_\I$ such that $\beta' \in
\sem{\bool{\maxnf}{\phi}}$. By the point
(\ref{it1:predicate-boolean}), $\I_{\beta'} \in \sem{\phi}$ and,
moreover, $\I_{\beta'} \subsetneq \I$, which contradicts the
minimality of $\I$. Thus $\beta_\I \in
\minsem{\bool{\maxnf}{\phi}}$. The other direction is symmetric.
\qed}

Considering the \mil\ formula
$\init{\asys} \isdef \bigvee_{k=1}^n \exists i_k
~.~ \typeno{\initstate}{k}(i_k)$, that defines the set of initial
configurations of a parametric system $\asys$, the following lemma
formalizes the intuition behind the definition of 
parametric trap constraints: 

\begin{lemma}\label{lemma:param-trap-constraint}
Let $\asys$ be a parametric system with interaction formula
  $\interform$ and $\I$ be a structure. Then
  $\I \models \trapconstraint{\interform} \wedge \init{\asys}$ iff
  $\set{(s,k) \mid k \in \iota(s)}$ is a marked trap of
  $\amarkednet_{\asys}$. Moreover,
  $\I \in \minsem{\trapconstraint{\interform} \wedge \init{\asys}}$
  iff $\set{(s,k) \mid k \in \iota(s)}$ is a minimal marked trap of
  $\amarkednet_{\asys}$.
\end{lemma}
\proof{
Let $\typeno{\acomptype}{k} = \tuple{\typeno{\ports}{k},
  \typeno{\states}{k}, \typeno{\initstate}{k}, \typeno{\rules}{k}}$
and define the bounded system: 
\[\begin{array}{rcl}
\unfolding{\asys} & \isdef & \tuple{ \set{ \tuple{\typeno{\ports}{k} \times \set{i}, 
\typeno{\states}{k} \times \set{i}, \typeno{\initstate}{k} \times \set{i}, 
\set{(s,i) \arrow{(p,i)}{} (s',i) \mid s \arrow{p}{} s' \in \typeno{\rules}{k}}} }_{
    \begin{array}{l}
      \scriptscriptstyle{k \in [1,n]} \\[-2mm]
      \scriptscriptstyle{i \in [1,\maxn{k}]}
  \end{array}}\!\!\!\!\!\!\!\!, \bool{\maxnf}{\interform} }
\end{array}\]
It is not hard to prove that $\amarkednet_\asys$ is the same as
$\amarkednet_{\unfolding{\asys}}$, thus their marked traps
coincide. The following equivalences follow from Lemma
\ref{lemma:predicate-boolean}:
\[\begin{array}{rcl}
\I \models \trapconstraint{\interform} & \iff & \beta_\I \models
\trapconstraint{\bool{\maxnf}{\interform}} \\ 
\I \models \init{\asys} & \iff & \beta_\I \models \init{\unfolding{\asys}}
\end{array}\]
Moreover, \(\set{(s,k) \mid k \in \iota(s)}
  = \set{(s,k) \mid \beta_\I((s,k)) = \true}\) and we apply Lemma
\ref{lemma:static-trap-constraint}. \qed
}

We are currently left with the task of computing a \mil\ formula which
defines the trap invariant $\alltrap{\amarkednet_\asys}$ of a
parametric component-based system
$\asys=\tuple{\typeno{\acomptype}{1}, \ldots, \typeno{\acomptype}{n},
  \maxnf, \interform}$. The difficulty lies in the fact that the size
of $\amarkednet_\asys$ and thus, that of the boolean formula
$\alltrap{\amarkednet_\asys}$ depends on the number $\maxn{k}$ of
instances of each component type $k \in [1,n]$. As we aim at computing
an invariant able to prove safety properties, such as deadlock
freedom, independently of how many components are present in the
system, we must define the trap invariant using a formula depending
exclusively on $\interform$, i.e.\ not on $\maxnf$.

Observe first that $\alltrap{\amarkednet_\asys}$ can be equivalently
defined using only the minimal marked traps of $\amarkednet_\asys$,
which, by Lemma \ref{lemma:param-trap-constraint}, are exactly the
sets $\set{(s,k) \mid k \in \iota(s)}$, defined by some structure
$(\U,\nu,\iota) \in \minsem{\trapconstraint{\interform} \wedge \init{\asys}}$.
Assuming that the set of structures
$\minsem{\trapconstraint{\interform} \wedge \init{\asys}}$, or an
over-approximation of it, can be defined by a positive \mil\ formula,
the trap invariant is defined using a generalization of boolean
dualisation to predicate logic, defined recursively, as follows:
\[\begin{array}{rclcrclcrclcrcl}
\dual{i=j} & \isdef & \neg i=j 
&\hspace*{.2cm}& 
\dual{\phi_1 \vee \phi_2} & \isdef & \dualnopar{\phi_1} \wedge \dualnopar{\phi_2} 
&\hspace*{.2cm}& 
\dual{\exists i~.~ \phi_1} & \isdef & \forall i~.~ \dualnopar{\phi_1}
&\hspace*{.2cm}& \dualnopar{p(i)} & \isdef & p(i) \\
\dual{\neg i=j} & \isdef & i=j 
&& 
\dual{\phi_1 \wedge \phi_2} & \isdef & \dualnopar{\phi_1} \vee \dualnopar{\phi_2} 
&& 
\dual{\forall i ~.~ \phi_1} & \isdef & \exists i ~.~ \dualnopar{\phi_1} 
\end{array}\]
The crux of the method is the ability of defining, given an arbitrary
\mil\  formula $\phi$, a positive \mil\  formula
$\ppos{\phi}$ that preserve its minimal models, formally $\phi
\minequiv \ppos{\phi}$. Because of quantification over unbounded domains, a \mil\ 
formula $\phi$ does not have a disjunctive normal form and thus one
cannot define $\ppos{\phi}$ by simply deleting the negative literals
in DNF, as was done for the definition of the positivation operation
$\pos{(.)}$, in the propositional case. For now we assume that the
transformation $\ppos{(.)}$ of monadic predicate formulae into
positive formulae preserving minimal models is defined (a detailed
presentation of this step is given next in
\S\ref{sec:cardinality}) and close this section with a parametric
counterpart of Theorem \ref{thm:static-trap-inv}.

\ifLongVersion
Before giving the proof of the main result of this section, we shall
be needing a few technical lemmas. For a set $\mathcal{S}$ of boolean
valuations, let
$\closure{\mathcal{S}} \isdef \set{\beta \mid \exists \beta' \in \mathcal{S}
~.~ \beta' \subseteq \beta}$ be its \emph{upward closure}. A set
$\mathcal{S}$ of boolean valuations is upward-closed iff
$\mathcal{S}=\closure{\mathcal{S}}$. The following lemma shows that
the set of models of a positive boolean formula is upward-closed and
thus uniquely determined by its minimal elements.

\begin{lemma}\label{lemma:pos-cl}
  Given a positive boolean formula $f$, we have $\sem{f}
  = \closure{\sem{f}} = \closure{\left(\minsem{f}\right)}$.
\end{lemma}
\proof{The inclusions
$\sem{f} \subseteq \closure{\sem{f}} \subseteq \closure{\left(\minsem{f}\right)}$
are immediate. To show that
$\closure{\left(\minsem{f}\right)} \subseteq \sem{f}$, observe that,
if $f$ is positive and $\beta \models f$ then any valuation $\beta'$
such that $\beta \subseteq \beta'$ is also model of $f$. Let
$\beta \in \closure{\minsem{f}}$ be a valuation. Then there exists
$\beta' \in \minsem{f}$ such that $\beta' \subseteq \beta$. Since
$\beta' \models f$ and $f$ is positive, we obtain $\beta \in
\sem{f}$. \qed}

\begin{lemma}\label{lemma:ppos}
  Given a \mil\  sentence $\phi$ with quantified variables
  $\typeno{i}{1}, \ldots, \typeno{i}{n}$ and a function $\maxnf :
  [1,n] \rightarrow \nat$, we have $\pos{\bool{\maxnf}{\phi}} \equiv
  \bool{\maxnf}{\ppos{\phi}}$.
\end{lemma}
\proof{It is sufficient to show $\pos{\bool{\maxnf}{\phi}}
\minequiv \bool{\maxnf}{\ppos{\phi}}$ and apply Lemma
\ref{lemma:pos-cl} to obtain the equivalence in general. To prove
$\minsem{\pos{\bool{\maxnf}{\phi}}} =
\minsem{\bool{\maxnf}{\ppos{\phi}}}$, we show that
$\minsem{\pos{\bool{\maxnf}{\phi}}} \subseteq
\sem{\bool{\maxnf}{\ppos{\phi}}}$ and
$\minsem{\bool{\maxnf}{\ppos{\phi}}} \subseteq
\sem{\pos{\bool{\maxnf}{\phi}}}$, respectively.

\noindent $\boxed{\minsem{\pos{\bool{\maxnf}{\phi}}} \subseteq
  \sem{\bool{\maxnf}{\ppos{\phi}}}}$ Let $\beta \in
\minsem{\pos{\bool{\maxnf}{\phi}}}$ be a valuation. Then, we also have
$\beta \in \minsem{\bool{\maxnf}{\phi}}$, since $\pos{\varphi}
\minequiv \varphi$, in general for any boolean formula
$\varphi$. Then, by Lemma \ref{lemma:predicate-boolean}
(\ref{it2:predicate-boolean}), there exists a structure $\I \in
\minsem{\phi}$ such that $\beta = \beta_\I$. Hence we obtain $\I \in
\minsem{\ppos{\phi}} \subseteq \sem{\ppos{\phi}}$. But then $\beta \in
\sem{\bool{\maxnf}{\ppos{\phi}}}$, by Lemma
\ref{lemma:predicate-boolean} (\ref{it1:predicate-boolean}).

\noindent $\boxed{\minsem{\bool{\maxnf}{\ppos{\phi}}} \subseteq
  \sem{\pos{\bool{\maxnf}{\phi}}}}$ Let $\beta \in
\minsem{\bool{\maxnf}{\ppos{\phi}}}$ be a boolean valuation. By Lemma
\ref{lemma:predicate-boolean} (\ref{it2:predicate-boolean}), we obtain
a structure $\I \in \minsem{\ppos{\phi}}$ such that $\beta =
\beta_\I$. But then $\I \in \minsem{\phi}$ and $\beta \in
\minsem{\bool{\maxnf}{\phi}}$, by Lemma \ref{lemma:predicate-boolean}
(\ref{it2:predicate-boolean}). Hence $\beta \in
\sem{\pos{\bool{\maxnf}{\phi}}}$. \qed}
\fi

\begin{theorem}\label{thm:parametric-trap-inv}
  For any parametric system $\asys
    = \tuple{\typeno{\acomptype}{1}, \ldots, \typeno{\acomptype}{n}, \maxnf, \interform}$,
    we
    have \[\begin{array}{c}\alltrap{\amarkednet_\asys} \equiv \bool{\maxnf}{\dual{\ppos{(\trapconstraint{\interform} \wedge \init{\asys})}}}\end{array}\]
\end{theorem}
\proof{
By Theorem \ref{thm:static-trap-inv}, we have
  $\alltrap{\amarkednet_\asys} \equiv
  \dual{\pos{\left(\trapconstraint{\bool{\maxnf}{\interform}} \wedge
        \bool{\maxnf}{\init{\asys}}\right)}}$. We obtain the following equivalences:
\[\begin{array}{rrl}
& \dual{\pos{\left(\trapconstraint{\bool{\maxnf}{\interform}} \wedge
        \bool{\maxnf}{\init{\asys}}\right)}} 
& \text{ since $\trapconstraint{\bool{\maxnf}{\interform}} \equiv \bool{\maxnf}{\trapconstraint{\interform}}$} \\
\equiv &  \dual{\pos{\left(\bool{\maxnf}{\trapconstraint{\interform}} \wedge
    \bool{\maxnf}{\init{\asys}}\right)}} \\ 
\equiv & \dual{\pos{\left(\bool{\maxnf}{\trapconstraint{\interform} \wedge
        \init{\asys}}\right)}} & \text{by Lemma \ref{lemma:ppos}} \\
\equiv & \dual{\bool{\maxnf}{\ppos{(\trapconstraint{\interform} \wedge
      \init{\asys})}}} \\
\equiv & \bool{\maxnf}{\dual{\ppos{\trapconstraint{\interform} \wedge
        \init{\asys}}}} & 
\end{array}\]\qed}

\section{Cardinality Constraints}
\label{sec:cardinality}

This section is concerned with the definition of a \emph{positivation}
operator $\ppos{(.)}$ for \mil\ sentences, whose only requirements are
that $\ppos{\phi}$ is positive and $\phi \minequiv \ppos{\phi}$. For
this purpose, we use a logic of quantifier-free \emph{boolean
cardinality constraints} \cite{KuncakNR06,BansalRBT16} as an
intermediate language, on which the positive formulae are defined. The
translation of \mil\ into cardinality constraints is done by an
equivalence-preserving quantifier elimination procedure, described
in \S\ref{sec:qe}. As a byproduct, since the satisfiability of
quantifier-free cardinality constraints is \np-complete
\cite{KuncakNR06} and integrated with SMT \cite{BansalRBT16}, we
obtain a practical decision procedure for \mil\ that does not use
model enumeration, as suggested by the small model property
\cite{Lowenheim15}. Finally, the definition of a positive
\mil\ formula from a boolean combination of quantifier-free
cardinality constraints is given in \S\ref{sec:pos}.

We start by giving the definition of cardinality constraints. Given
the set of monadic predicate symbols $\preds$, a \emph{boolean term}
is generated by the syntax:
\[t := \apred \in \preds \mid \neg t_1 \mid t_1 \wedge t_2 \mid t_1 \vee t_2\]
When there is no risk of confusion, we borrow the terminology of
propositional logic and say that a term is in DNF if it is a
disjunction of conjunctions (minterms). We also write $t_1 \rightarrow
t_2$ if and only if the implication is valid when $t_1$ and $t_2$ are
interpreted as boolean formulae, with each predicate symbol viewed as
a propositional variable. Two boolean terms $t_1$ and $t_2$ are said
to be \emph{compatible} if and only if $t_1 \wedge t_2$ is
satisfiable, when viewed as a boolean formula.

For a boolean term $t$ and a first-order variable $i \in \vars$, we
define the shorthand $t(i)$ recursively, as \((\neg t_1)(i) \isdef
\neg t_1(i)\), \((t_1 \wedge t_2)(i) \isdef t_1(i) \wedge t_2(i)\) and
\((t_1 \vee t_2)(i) \isdef t_1(i) \vee t_2(i)\). Given a positive integer 
$n \in \nat$ and $t$ a boolean term, we define the following
\emph{cardinality constraints}, by \mil\ formulae:
\[\begin{array}{rclcrcl}
\len{t} \geq n & \isdef & \exists i_1 \ldots \exists i_n ~.~
\distinct(i_1,\ldots,i_n) \wedge \bigwedge_{j=1}^n t(i_j) & \hspace*{1cm} &
\len{t} \leq n & \isdef & \neg(\len{t} \geq n+1) 
% \len{t} = n \isdef \len{t} \geq n \wedge \len{t} \leq n
\end{array}\]
We shall further use cardinality constraints with $n = \infty$, by
defining $\len{t} \geq \infty \isdef \false$ and
$\len{t} \leq \infty \isdef \true$. The intuitive semantics of
cardinality constraints is formally defined in terms of structures
$\I=(\U,\nu,\iota)$ by the semantics of monadic predicate logic, given
in the previous. For instance, $\len{p\wedge q}
\geq 1$ means that the intersection of the sets $p$ and $q$ is not
empty, whereas $\len{\neg p} \leq 0$ means that $p$ contains all
elements from the universe.

\subsection{Quantifier Elimination}
\label{sec:qe}

Given a sentence $\phi$, written in \mil, we build an equivalent
boolean combination of cardinality constraints $\qelim{\phi}$, using
quantifier elimination. We describe the elimination of a single
existential quantifier and the generalization to several existential
or universal quantifiers is immediate. Assume that $\phi = \exists i_1
~.~ \bigvee_{k \in K}
\psi_k(i_1,\ldots,i_m)$, where $K$ is a finite set of indices and, for
each $k\in K$, $\psi_k$ is a quantifier-free conjunction of atomic
propositions of the form $i_j=i_\ell$, $\apred(i_j)$ and their negations,
for some $j,\ell \in [1,m]$. We write, equivalently, $\phi \equiv \bigvee_{k \in K}
\varphi_k \wedge \exists i_1 ~.~ \theta_k(i_1,\ldots,i_m)$, where
$\varphi_k$ does not contain occurrences of $i_1$ and $\theta_k$ is a
conjunction of literals of the form $\apred(i_1)$, $\neg \apred(i_1)$,
$i_1 = i_j$ and $\neg i_1 = i_j$, for some $j \in [2,m]$. For each
$k \in K$, we distinguish the following cases: \begin{compactenum}
\item if $i_1=i_j$ is a consequence of $\theta_k$, for some $j>1$, 
let $\qelim{\exists i_1 ~.~ \theta_k} \isdef \theta_k[i_j/i_1]$.
\item else, \(\theta_k = \bigwedge_{j \in J_k} \neg i_1 = i_j \wedge t_k(i_1)\)
for some $J_k \subseteq [2,m]$ and boolean term $t_k$, and let:
\[\begin{array}{rcl}
\qelim{\exists i_1 ~.~ \theta_k} & \isdef & \bigwedge_{J \subseteq J_k} 
\Big[\distinct\big(\{i_j\}_{j\in J}\big) \wedge \bigwedge_{j \in J} t_k(i_j)\Big] 
\rightarrow \len{t_k} \geq \card{J}+1 \\
\qelim{\phi} & \isdef & \bigvee_{k \in K} \varphi_k \wedge \qelim{\exists i_1 ~.~ \theta_k}
\end{array}\]
\end{compactenum}
Universal quantification is dealt with using the duality
$\qelim{\forall i_1 ~.~ \psi} \isdef \neg\qelim{\exists i_1
~.~ \neg\psi}$. For a prenex formula $\phi = Q_n i_n \ldots Q_1 i_1
~.~ \psi$, where $Q_1, \ldots, Q_n \in \set{\exists,\forall}$ and
$\psi$ is quantifier-free, we define, recursively
$\qelim{\phi} \isdef \qelim{Q_n i_n ~.~ \qelim{Q_{n-1} i_{n-1} \ldots
Q_1 i_1 ~.~ \psi}}$. It is easy to see that, if $\phi$ is a sentence,
$\qelim{\phi}$ is a boolean combination of cardinality
constraints. The correctness of the construction is a consequence of
the following lemma:

\begin{lemma}\label{lemma:qe-equiv}
  Given a \mil\ formula $\phi=Q_n i_n \ldots Q_i i_1 ~.~ \psi$, where
  $Q_1, \ldots, Q_n \in \set{\forall,\exists}$ and $\psi$ is a
  quantifier-free conjunction of equality and predicate atoms, we have
  $\phi\equiv\qelim{\phi}$.
\end{lemma}
\proof{
  We give the proof only for the case $n=1$ and $Q_1=\exists$, the
  general case being an easy consequence. Suppose that $\psi
  = \varphi \wedge \theta(i_1)$, where $i_1$ does not occur within
  $\varphi$. If $\theta \models i_1 = i_j$ for some $j \neq 1$ then
  $\exists i_1 ~.~ \theta \equiv \theta[i_j/i_1]$. Otherwise, let
  $\theta = \bigwedge_{j \in J} \neg i_1 = i_j \wedge t_j(i_1)$, for
  some boolean terms $t_j$ and show: \[\exists i_1
  ~.~ \theta \equiv \bigwedge_{K \subseteq
  J} \left(\distinct(\set{i_k}_{k \in K}) \wedge \bigwedge_{k \in K}
  t(i_k)\right) \rightarrow \len{t} \geq \card{K}+1\]

  \noindent ``$\Rightarrow$'' Let $(\U,\nu[i_1
    \leftarrow u],\iota) \models \bigwedge_{j \in J} \neg i_1 = i_j
  \wedge t(i_1)$, for some $u \in \U$ and let $K$ be the maximal
  subset of $J$ such that $\nu(i_{k_1}) \neq \nu(i_{k_2})$, for all
  $k_1 \neq k_2 \in K$ and $\nu(i_j) \in \iota(t)$. Since, moreover,
  $\nu(i_1) \not\in \set{\nu(i_k)}_{k \in k}$, we obtain
  $\card{\iota(t)} \geq \card{K}+1$.

  \vspace*{\baselineskip}\noindent ``$\Leftarrow$'' Let
  $(\U,\nu,\iota)$ be a model of the right-hand side formula and let
  $K \subseteq J$ be a set such that $\nu(x_{k_1}) \neq \nu(x_{k_2})$
  for all $k_1 \neq k_2 \in K$ and $\set{\nu(x_k)}_{k \in K} \in
  \iota(t)$. Then, since $\card{\iota(t)} \geq \card{K}+1$, there
  exists $u \in \iota(t) \setminus \set{\nu(x_k)}_{k \in K}$ and thus
  $(\U,\nu[i_1 \leftarrow u],\iota) \models \bigwedge_{j \in J} \neg
  x_1 = x_j \wedge t(x_1)$. \qed}

\begin{example}\label{ex:qe}
(contd. from Example \ref{ex:tc})
% \comment[ri]{q.e. for the trap constraint + init condition for Fig. \ref{fig:ex-intro}}
%| [ not(r(1)) le([w,not(u)],0) not(s(1)) le([u,not(w)],0) le(1,[w]) ]
%| [ not(r(1)) le([w,not(u)],0) le([not(w)],0) le(1,[w]) ]
%| [ s(1) r(1) ]
%| [ s(1) le([not(w)],0) le(1,[w]) ]
%| [ le([not(u)],0) not(s(1)) le([u,not(w)],0) le(1,[w]) ]
%| [ le([not(u)],0) r(1) ]
%| [ le([not(u)],0) le([not(w)],0) le(1,[w]) ]
Below we show the results of quantifier elimination applied to the
conjunction $\trapconstraint{\interform} \wedge \init{\asys}$
for the system in Figure \ref{fig:ex-intro-parametric}: 
\[\begin{array}{c}
(\neg r \wedge \neg s \wedge |w \wedge \neg u| \le 0 \wedge |u \wedge \neg w|\le 0 \wedge 1 \le |w|) ~\vee \\
(\neg r \wedge |w \wedge \neg u| \le 0 \wedge |\neg w|\le 0 \wedge 1 \le |w|) \vee 
(s \wedge r) \vee (s \wedge |\neg w| \le 0 \wedge 1 \le |w|) ~\vee \\
(\neg s \wedge |\neg u| \le 0 \wedge |u \wedge \neg w| \le 0 \wedge 1 \le |w|) \vee
(|\neg u| \le 0 \wedge |\neg w| \le 0 \wedge 1 \le |w|) \enspace. 
\end{array}\]
Similarly, for the system in Figure \ref{fig:bip-broadcast}, we obtain
the following cardinality constraints:
% \comment[ri]{the one from Fig. \ref{fig:bip-broadcast}}
%| [ le(3,[w]) le([u,not(w)],0) ]
%| [ le(2,[w]) le([w,not(u)],1) le([u,not(w)],0) ]
%| [ le([not(u)],1) le([not(u),not(w)],0) le([u,not(w)],0) le(1,[w]) ]
%| [ le([w,not(u)],0) le([u,not(w)],0) le(1,[w]) ]
\[ \begin{array}{c}
  (3 \le |w| \wedge |u \wedge \neg w| \le 0) \vee 
  (2 \le |w| \wedge |w \wedge \neg u| \le 1 \wedge |u \wedge \neg w| \le 0) ~\vee \\
  (|\neg u| \le 1 \wedge |\neg u \wedge \neg w| \le 0 \wedge |u \wedge \neg w| \le 0 \wedge 1 \le |w|) \vee
  (|w \wedge \neg u| \le 0 \wedge |u \wedge \neg w| \le 0 \wedge 1 \le |w|)\enspace. 
\end{array}\]
\end{example}

\subsection{Building Positive Formulae that Preserve Minimal Models}
\label{sec:pos}

Let $\phi$ be a \mil\ formula, not necessarily positive. We shall
build a positive formula $\ppos{\phi}$, such that
$\phi \minequiv \ppos{\phi}$.  By the result of the last section,
$\phi$ is equivalent to a boolean combination of cardinality
constraints $\qelim{\phi}$, obtained by quantifier elimination.  Thus
we assume w.l.o.g. that the DNF of $\phi$ is a disjunction of
conjunctions of the form $\bigwedge_{i \in
L} \len{t_i} \geq \ell_i \wedge \bigwedge_{j \in U} \len{t_j} \leq
u_j$, for some sets of indices $L$, $U$ and some positive integers
$\{\ell_i\}_{i \in L}$ and $\{u_j\}_{j \in U}$.

For a boolean combination of cardinality constraints $\psi$, we denote
by $\voc{\psi}$ the set of predicate symbols that occur in a boolean
term of $\psi$ and by $\posvoc{\psi}$ ($\negvoc{\psi}$) the set of
predicate symbols that occur under an even (odd) number of negations
in $\psi$. The following proposition allows to restrict the form of
$\phi$ even further, without losing generality:

\begin{proposition}\label{prop:minsem-distr}
Given \mil\ formulae $\phi_1$ and $\phi_2$, for any positivation
operator $\ppos{(.)}$, the following hold: \begin{compactenum}
\item\label{it1:minsem-distr} $\ppos{(\phi_1 \vee \phi_2)} \minequiv \ppos{\phi_1} \vee \ppos{\phi_2}$, 
\item\label{it2:minsem-distr} $\ppos{(\phi_1 \wedge \phi_2)} \minequiv \ppos{\phi_1} \wedge \ppos{\phi_2}$, 
provided that $\voc{\phi_1} \cap \voc{\phi_2} = \emptyset$.
\end{compactenum}
\end{proposition}
From now on, we assume that $\phi$ is a conjunction of cardinality
constraints that cannot be split as $\phi = \phi_1 \wedge \phi_2$,
such that $\voc{\phi_1} \cap \voc{\phi_2} = \emptyset$.

Let us consider a cardinality constraint $\len{t} \geq \ell$ that
occurs in $\phi$. Given a set $\mathcal{P}$ of predicate symbols, for
a set of predicates $S \subseteq \mathcal{P}$, the
\emph{complete} boolean minterm corresponding to $S$ with respect to
$\mathcal{P}$ is \(\compt{S}{\mathcal{P}} \isdef \bigwedge_{p \in S}
p \wedge \bigwedge_{p \in \mathcal{P} \setminus S} \neg p\).
Moreover, let $\mathcal{S}_t \isdef \set{S \subseteq \voc{\phi} \mid
t_S \rightarrow t}$ be the set of sets $S$ of predicate symbols for
which the complete minterm $t_S$ implies $t$. Finally, each
cardinality constraint $\len{t} \geq \ell$ is replaced by the
equivalent disjunction\footnote{The constraints $\len{t} \leq u$ are
dealt with as $\neg(\len{t} \geq u+1)$.}, in which each boolean term
is complete with respect to $\voc{\phi}$:
\[\len{t} \geq \ell \equiv \bigvee \Big\{\bigwedge_{S \in \mathcal{S}_t} \len{\compt{S}{\voc{\phi}}}
\geq \ell_S \mid \text{ for some constants $\set{\ell_S \in \nat}_{S \in \mathcal{S}_t}$ such that }
\sum_{S \in \mathcal{S}_t} \ell_S = \ell\Big\}\] Note that because any 
two complete minterms $t_S$ and $t_T$, for $S \neq T$, are
incompatible, then necessarily $\len{t_S \vee t_T} = \len{t_S}
+ \len{t_T}$. Thus $\len{t_S \vee t_T} \geq \ell$ if and only if there
exist $\ell_1, \ell_2 \in \nat$ such that $\ell_1+\ell_2 = \ell$ and
$\len{t_S} \geq \ell_1$, $\len{t_T} \geq \ell_2$, respectively.

Notice that, restricting the sets of predicates in $\mathcal{S}_t$ to
subsets of $\voc{\phi}$, instead of the entire set of predicates,
allows to apply Proposition \ref{prop:minsem-distr} and reduce the
number of complete minterm to be considered. That is, whenever
possible, we write each minterm $\bigwedge_{i \in
L} \len{t_i} \geq \ell_i \wedge \bigwedge_{j \in U} \len{t_j} \leq
u_j$ in the DNF of $\phi$ as $\psi_1 \wedge \ldots \wedge \psi_k$,
such that $\voc{\psi_i} \cap \voc{\psi_j} = \emptyset$ for all $1 \leq
i < j \leq k$. In practice, this optimisation turns out to be quite
effective, as shown by the small execution times of our test cases,
reported in \S\ref{sec:experiments}.

The second step is building, for each conjunction $C
= \bigwedge \set{\ell_S \leq \len{\compt{S}{\voc{\phi}}} \wedge \len{\compt{S}{\voc{\phi}}} \leq
u_S \mid S \subseteq \voc{\phi}}$\footnote{Missing lower bounds
$\ell_S$ are replaced with $0$ and missing upper bounds $u_S$ with
$\infty$.}, as above, a positive formula $\ppos{C}$, that preserves
its set of minimal models $\minsem{C}$. The generalization to
arbitrary boolean combinations of cardinality constraints is a direct
consequence of Proposition \ref{prop:minsem-distr}. Let
$\poslan{\phi}$ (resp. $\neglan{\phi}$) be the set of positive boolean
combinations of predicate symbols $p \in \posvoc{\phi}$ (resp. $\neg
p$, where $p \in \negvoc{\phi}$). Further, for a complete minterm
$\compt{S}{\mathcal{P}}$, we write ${\compt{S}{\mathcal{P}}}^+$
(${\compt{S}{\mathcal{P}}}^-$) for the conjunction of the positive
(negative) literals in $\compt{S}{\mathcal{P}}$. Then, we define:
\[\begin{array}{c}
\ppos{C} \isdef \bigwedge \Big\{\len{\tau} \geq \sum_{{\compt{S}{\voc{\phi}}}^+ \rightarrow \tau} \ell_S \mid \tau \in \poslan{\phi}\Big\} \wedge \bigwedge \Big\{\len{\tau} \leq \sum_{{\compt{S}{\voc{\phi}}}^- \rightarrow \tau} u_S \mid \tau \in \neglan{\phi}\}
\end{array}\]
It is not hard to see that $\ppos{C}$ is a positive \mil\ formula,
because: \begin{compactitem}
\item for each $\tau \in \poslan{\phi}$, we have 
      $\len{\tau} \geq k \equiv \exists i_1 \ldots \exists i_k
      ~.~ \distinct(i_1,\ldots,i_k) \wedge \bigwedge_{j=1}^k \tau(j)$ and
\item for each $\tau \in \neglan{\phi}$, we have
      $\len{\tau} \leq k \equiv \forall i_1 \ldots \forall i_{k+1} ~.~ \distinct(i_1,\ldots,i_{k+1}) \rightarrow \bigvee_{j=1}^{k+1} \neg\tau(i_j)$.
\end{compactitem}
The following lemma proves that the above definition meets the second
requirement of positivation operators, concerning the preservation of
minimal models.

\begin{lemma}\label{lemma:positivation}
Given $\mathcal{P}$ a finite set of monadic predicate symbols,
$\set{\ell_S \in \nat}_{S \subseteq \mathcal{P}}$ and
$\set{u_S \in \nat \cup \set{\infty}}_{S \subseteq \mathcal{P}}$ sets
of constants, for any conjunction $C
= \bigwedge \set{\ell_S \leq \len{\compt{S}{\mathcal{P}}} \wedge \len{\compt{S}{\mathcal{P}}} \leq
u_S \mid S \subseteq \mathcal{P}}$, we have $C \minequiv \ppos{C}$.
\end{lemma}

\begin{example}\label{ex:pos}(contd. from Example \ref{ex:qe})

% \comment[mb]{positive forms of the two formulae in Example \ref{ex:qe} - 1st conjuct of first example}

Consider the first minterm of the DNF of the cardinality constraint
obtained by quantifier elimination in Example \ref{ex:qe}, from the
system in Figure \ref{fig:ex-intro-parametric}. The result of
positivation for this minterm is given below:
\[\ppos{\big(\neg r \wedge \neg s \wedge |w \wedge \neg u| \le 0 
\wedge |u \wedge \neg w|\le 0 \wedge 1 \le |w|\big)}= 1 \le |u \wedge w|\]
Intuitively, the negative literals $\neg r$ and $\neg s$ may safely
disapear, because no minimal model will assign $r$ or $s$ to
true. Further, the constraints $\len{w \wedge \neg u} \le 0$ and
$\len{u \wedge \neg w} \leq 0$ are equivalent to the fact that, in any
structure $\I = (\U, \nu, \iota)$, we must have
$\iota(u)=\iota(w)$. Finally, because $\len{w}\ge1$, then necessarily
$\len{u \wedge w} \ge 1$.

% \comment[mb]{positive forms of the two formulae in Example \ref{ex:qe} - 2nd conjuct of second example}

Similarly, the result of positivation applied to the second conjunct
of the DNF cardinality constraint corresponding to the system in
Figure \ref{fig:bip-broadcast} is given below:
\[\ppos{\big(2 \le |w| \wedge |w \wedge \neg u| \le 1 \wedge |u \wedge \neg w| \le 0\big)} = 
2 \le |w| \wedge 1 \le |u \wedge w|\] Here, the number of elements in
$w$ is at least $2$ and, in any structure $\I = (\U, \nu, \iota)$, we
must have $\iota(u) \subseteq \iota(w)$ and at most one element in
$\iota(w) \setminus \iota(u)$. Consequently, the intersection of the
sets $\iota(u)$ and $\iota(w)$ must contain at least one element,
i.e.\ $\len{u \wedge w} \geq 1$. 
\end{example}

\ifLongVersion
\subsubsection{The Proof of Lemma \ref{lemma:positivation}}

This is the most intricate technical result of the paper, that
requires several additional notions, which are the concern of this
section. If $t$ is any boolean term, its interpretation in the
structure $\I=(\U,\nu,\iota)$ is the set $t^\I \subseteq \U$ defined
recursively, as follows:
\[
p^\I \isdef \iota(p) \hspace*{1cm} (\neg t)^\I \isdef \U \setminus t^\I \hspace*{1cm}
(t_1 \wedge t_2)^\I \isdef t_1^\I \cap t_2^\I \hspace*{1cm} (t_1 \vee t_2)^\I \isdef t_1^\I \cup t_2^\I
\]
Next, we generalize upward closures and upward closed sets
from boolean valuations to first order structures as follows.
If $\mathcal{S}$ is a set of structures sharing the same universe,
then
$\upclose{\mathcal{S}} \isdef \set{\I \mid \exists \I' \in \mathcal{S}
~.~ \I' \subseteq \I}$ denotes its upward closure. Moreover,
$\mathcal{S}$ is upward closed iff $\mathcal{S}
= \upclose{\mathcal{S}}$. Then we have the following facts, whose
proofs are folklore:

\begin{fact}\label{fact:upclosed}
Given a positive \mil\ formula $\phi$, the set $\sem{\phi}$ is upward
closed.
\end{fact}
\proof{By induction on the structure of $\phi$. \qed}

\begin{fact}\label{fact:mineq}
Given \mil\ formulae $\phi_1$ and $\phi_2$, we have
$\phi_1 \minequiv \phi_2$ if and only if $\upclose{\sem{\phi_1}}
= \upclose{\sem{\phi_2}}$. If, moreover, $\phi_2$ is positive,
$\phi_1 \minequiv \phi_2$ if and only if $\upclose{\sem{\phi_1}}
= \sem{\phi_2}$.
\end{fact}
\proof{The first point is due to the observation 
$\upclose{\sem{\phi_i}} = \set{\I \mid \exists \I' \in \minsem{\phi_i}
~.~ \I' \subseteq \I}$, for $i=1,2$. The second point is obtained
applying Fact \ref{fact:upclosed}. \qed}

Given a conjunction $C = \bigwedge \set{\ell_S \leq \len{\compt{S}{\mathcal{P}}}
\wedge \len{\compt{S}{\mathcal{P}}} 
\leq u_S \mid S \subseteq \mathcal{P}}$ of cardinality constraints involving
all complete minterms with respect to $\mathcal{P}$, for some
arbitrary \mil\ formula $\phi$, Lemma \ref{lemma:positivation}
requires showing that $\upclose{\sem{C}}=\sem{\ppos{C}}$. We shall do
this in two stages: \begin{compactenum}
\item We express $\upclose{\sem{C}}$ using the reachability set of a vector 
addition system with states of a special form, that is, moreover,
definable as the set of solutions of an integer linear system.
\item We use Hoffman's Circulation Theorem \cite[Theorem 11.2]{Schrijver03} 
to show that the set of solutions of the linear system above defines
$\sem{\ppos{C}}$.
\end{compactenum} 
The developments of the two points rely on the observation that, each
model of a cardinality constraint is uniquely defined, up to the
renaming of its elements, by the positive cardinality of each complete
minterm. In the following we shall consider this fact implicit and
work with mappings of minterms into positive integer values, instead
of first order structures.

\vspace*{\baselineskip}
\paragraph{\em Vector Addition Systems}
The goal of this paragraph is to reduce the problem $\upclose{\sem{C}}
= \sem{\ppos{C}}$ of equivalence between sets of first order
structures to the resolution of a linear integer system. To begin with
observe that, given an arbitrary \mil\ formula $\phi$, if
$\I \models \phi$, then any structure obtained from $\I$ by a renaming
of its elements is also a model of $\phi$. This is because $\phi$ uses
only equalities and disequalities, which cannot distigush the
particular identity of elements. In other words, $\sem{\phi}$ is
closed under isomorphic transformations of structures.

In the following, we assume that the finite set $\mathcal{P}$ of
predicate symbols is indexed by a total order $\lhd$. Then any set
$S \subseteq \mathcal{P}$ corresponds to a word $w_S$ which is the
sequence of its elements, in the $\lhd$ order. Moreover, let
$\lhd_{\mathit{lex}}$ be the lexicographic order induced by $\lhd$.
The following definition introduces a total order on sets of predicate
symbols, that is compatible with the subset ordering.

\begin{definition}\label{def:set-order}
Given sets $S,T \subseteq \mathcal{P}$, where $\mathcal{P}$ is totally
ordered via $\lhd$, we define the total order $S \setord T$ if and
only if one of the following holds: \begin{compactenum}
\item $S \subseteq T$, or
\item $S \not\subseteq T$ and $w_S \lhd_{\mathit{lex}} w_T$. 
\end{compactenum}
\end{definition}
As usual, we write $S \setordneq T$ for $S \setord T$ and $S \neq T$. 

Let
$\vec{t}^\I \isdef \tuple{\card{(\compt{S}{\mathcal{P}})^\I}}_{S \subseteq \mathcal{P}}$
be the vector of cardinalities of the interpretations for all complete
minterms in the structure $\I$, arranged in the $\setord$ order. In
the following, we sometimes refer to this vector as
the \emph{cardinality vector} of the structure $\I$.

Since the interpretations of the complete minterms w.r.t $\mathcal{P}$
are pairwise disjoint, for any predicate symbol $p \in \mathcal{P}$,
we have $p^\I = \bigcup_{p \in S} (\compt{S}{\mathcal{P}})^\I$. Using
the recursive definitions above, we can write any boolean term as a
finite union of complete minterms, which corresponds to the DNF of the
boolean formula associated with it. Hence, for any cardinality
constraint $\len{t} \geq n$, we have $\I \models \len{t} \geq n$ if
and only if $\sum_{i=1}^k \card{(\compt{S_i}{\mathcal{P}})^I} \geq n$,
where $\compt{S_1}{\mathcal{P}},\ldots,\compt{S_k}{\mathcal{P}}$ is
the set of complete minterms that occur in the DNF of $t$. In general,
for a boolean combination of cardinality constraints $\phi$, we write
$\vec{t}^\I \models \phi$ if and only if the formula obtained by
replacing each term $\len{t}$ with the sum above is logically valid.
A formal definition can be given recursively, on the structure of
$\phi$.

At this point, we can identify the set of models $\sem{\phi}$, where
$\phi$ is any boolean combination of cardinality constraints, by the
set of vectors $\set{\vec{t}^\I \mid \vec{t}^\I \models \phi}$, up to
isomorphism of first order structures. It remains now to define upward
closures in the same way. A first remark is that, because the set
$\sem{\phi}$ is closed under isomorphism, so is its upward closure
$\upclose{\sem{\phi}}$. However, the definition of
$\upclose{\sem{\phi}}$ in terms of vectors $\vec{t}^\I$ requires a
partial order that captures the pointwise inclusion between structures
$\I \subseteq \I'$.

\begin{definition}\label{def:cardinality-po}
Given structures $\I$ and $\I'$ with the same universe, we
define the relation $\vec{t}^{\I'} \prec_1 \vec{t}^{\I}$ if and only
if there exists a set $S \subseteq \mathcal{P}$ and a predicate symbol
$p \in S$ such that: \begin{compactenum}
\item $\card{(\compt{S}{\mathcal{P}})^\I} = \card{(\compt{S}{\mathcal{P}})^{\I'}} + 1$,
\item $\card{(\compt{S\setminus\set{p}}{\mathcal{P}})^\I} = \card{(\compt{S\setminus\set{p}}{\mathcal{P}})^{\I'}} - 1$,
\item $\card{(\compt{T}{\mathcal{P}})^\I} = \card{(\compt{T}{\mathcal{P}})^{\I'}}$, for all
$T \subseteq \mathcal{P}$, such that $T \neq S$ and $T \neq S \setminus\set{p}$.
\end{compactenum}
We denote by $\preceq$ the reflexive and transitive closure of the
$\prec_1$ relation.
\end{definition}

\begin{lemma}\label{lemma:cardinality}
For a boolean combination of cardinality constraints $\phi$, the
following hold:
\begin{compactenum}
\item\label{it1:cardinality} $\sem{\phi} = \set{\I \mid \vec{t}^\I \models \phi}$, 
\item\label{it2:cardinality} $\upclose{\sem{\phi}} =
\set{\I \mid \exists \I' \in \sem{\phi} ~.~ \vec{t}^{\I'} \preceq \vec{t}^{\I}}$.
\end{compactenum}
\end{lemma}
\proof{
(\ref{it1:cardinality}) One shows that, for any structure $\I$, we
have $\I \models \phi \iff \vec{t}^\I \models \phi$, by induction on
the structure of $\phi$. The base case $\phi = \len{t}\geq n$ is by
definition and the inductive steps are
routine. (\ref{it2:cardinality}) We show that, for any structure
$\I=(\U,\nu,\iota)$, the following are
equivalent: \begin{compactenum}[(i)]
\item\label{it2:1:cardinality} there exists $\I' \in \sem{\phi}$ such that $\I' \subseteq \I$, and
\item\label{it2:2:cardinality} there exists $\I'' \in \sem{\phi}$ such that $\vec{t}^{\I''} \preceq \vec{t}^{\I}$.
\end{compactenum}
(\ref{it2:1:cardinality}) $\Rightarrow$ (\ref{it2:2:cardinality}) We
let $\I'' = \I'$ and prove $\vec{t}^{\I'} \preceq \vec{t}^{\I}$. If,
for all predicate symbols $p \in \mathcal{P}$, we have $p^{\I'} =
p^{\I}$, then $\I'=\I$ and $\vec{t}^{\I'} = \vec{t}^{\I}$ follows.
Assuming that this is not the case, let $p$ be an arbitrary predicate
symbol such that $p^{\I'} \subset p^{\I}$. We build a sequence of
structures $\I=\I_0, \ldots, \I_k$ such that
$p^{\I_0} \supset \ldots \supset p^{\I_k}$ and
$\vec{t}^{\I_0} \succ_1 \ldots \succ_1 \vec{t}^{\I_k}$.  Let $u \in
p^{\I} \setminus p^{\I'}$ be an element and let
$S_u\isdef\set{q \in \mathcal{P} \mid u \in q^{\I}}$. Clearly, we have
that $p \in S_u$. Let $\I_1=(\U,\nu,\iota_1)$ be the structure such
that $\iota_1(p)=\iota(p) \setminus \set{u}$ and $\iota_1(q)=\iota(q)$
for all $q \in \mathcal{P} \setminus \set{p}$. It is not hard to see
that: \begin{compactitem}
\item $\card{(\compt{S_u}{\mathcal{P}})^\I} = \card{(\compt{S_u}{\mathcal{P}})^{\I_1}} + 1$, 
\item $\card{(\compt{S_u\setminus\set{p}}{\mathcal{P}})^\I} = \card{(\compt{S_u\setminus\set{p}}{\mathcal{P}})^{\I_1}} - 1$, 
\item $\card{(\compt{T}{\mathcal{P}})^\I} = \card{(\compt{T}{\mathcal{P}})^{\I_1}}$,
for all $T \subseteq \mathcal{P}$, such that $T \neq S_u$ and $T \neq S_u \setminus \set{p}$.
\end{compactitem} 
By Definition \ref{def:cardinality-po}, we have
$\vec{t}^{\I_0} \succ_1 \vec{t}^{\I_1}$. We continue chosing elements
$u \in p^{\I} \setminus p^{\I'}$ until no such elements can be found,
then pick another predicate symbol for which $\I$ and $\I'$ differ. In
this way we obtain a finite sequence of structures
$\set{\I_j}_{j=0}^n$, such that $\I_j \succ_1 \I_{j+1}$ for all
$0 \leq j < n$, thus $\vec{t}^{\I} \succeq \vec{t}^{\I'}$, as required. 

\noindent (\ref{it2:2:cardinality}) $\Rightarrow$ (\ref{it2:1:cardinality})
By induction on the length of the sequence of structures
$\I=\I_0, \ldots, \I_k=\I''$ such that
$\vec{t}^{\I_0} \succ_1 \ldots \succ_1 \vec{t}^{\I_k}$. In the base
case $k=0$, we have $\vec{t}^{\I''} = \vec{t}^{\I}$, thus we have
$\vec{t}^{\I} \models \phi$ and consequently $\I \in \sem{\phi}$, by
point (\ref{it1:cardinality}). For the induction step $k>0$, we
observe that $\vec{t}^{\I_0} \succ_1 \vec{t}^{\I_1}$ implies the
existence of a structure $\I'_1 \subset \I_0$ which is isomorphic to
$\I_1$, thus $\vec{t}^{\I'_1} = \vec{t}^{\I_1}$. By the induction
hypothesis, there exists $I' \in \sem{\phi}$ such that
$\I' \subseteq \I'_1$, hence $\I' \subseteq \I$, as required. 
\qed}

In the following, we define a vector addition system whose
reachability relation matches the $\preceq$ partial order on
cardinality vectors $\vec{t}^\I$. 

\begin{definition}\label{def:vas}
An $n$-dimensional vector addition system (VAS) is a finite set of
vectors $V = \set{\vec{v}_1, \ldots, \vec{v}_k} \subseteq \zed^n$.
\end{definition}
A configuration of $V = \set{\vec{v}_1, \ldots, \vec{v}_k}$ is a
vector $\vec{c} \in \nat^n$. The one-step reachability relation in $V$
is $\vec{c} \arrow{\vec{v}_i}{V} \vec{c'}$ if and only if $\vec{c'}
= \vec{c} + \vec{v}_i$, for some $1 \leq i \leq k$. The fact that
$\vec{c}, \vec{c}' \in \nat^n$ is important here, because
configurations of a VAS are not allowed to contain negative values.
For a finite sequence $\sigma = \vec{v}_{i_1} \ldots \vec{v}_{i_k}$ of
vectors from $V$, we write $\vec{c} \arrow{\sigma}{V} \vec{c'}$ for
the sequence of transitions $\vec{c} \arrow{\vec{v}_{i_1}}{V} \vec{c}_1
\arrow{\vec{v}_{i_2}}{V} \ldots \arrow{\vec{v}_{i_k}}{V} \vec{c'}$. Moreover, 
we write $\vec{c} \arrow{*}{V_{\mathcal{P}}} \vec{c}'$ when $\sigma$ is not important. 

For a vector $\vec{v} \in \set{-1,0,1}^{2^{\card{\mathcal{P}}}}$ and a
set $S \subseteq \mathcal{P}$, let $\vec{v}(S)$ be the entry in
$\vec{v}$ corresponding to $S$. Moreover, for some predicate symbol
$p \in S$, we denote by $\vec{v}(S,p)$ the vector $\vec{u}$ such that
$\vec{u}(S) = -1$, $\vec{u}(S\setminus\set{p}) = 1$ and $\vec{u}(T) =
0$, for all $T \subseteq \mathcal{P}$ such that $T \neq S$ and $T \neq
S \setminus \set{p}$. Intuitively, $\vec{v}(S,p)$ transfers an element
from $\compt{S}{\mathcal{P}}$ into
$\compt{S\setminus\set{p}}{\mathcal{P}}$, thus decreasing the
cardinality of $\compt{S}{\mathcal{P}}$ and increasing that of
$\compt{S\setminus\set{p}}{\mathcal{P}}$ by one, respectively.

We now define the $2^{\card{\mathcal{P}}}$-dimensional VAS
$V_{\mathcal{P}} \isdef \set{\vec{v}(S,p) \mid
S \subseteq \mathcal{P}, p \in S}$.
%% A sequence of vectors
%% $\vec{v}(S_1,p_1), \ldots, \vec{v}(S_k,p_k) \in V_{\mathcal{P}}$
%% is \emph{ordered} if, for each $1 \leq i < k$,
%% either \begin{inparaenum}[(i)]
%% \item $S_{i+1} \lhd^\dagger S_i$, or
%% \item $S_{i+1} = S_{i}$ and $p_{i+1} \unlhd p_{i}$. 
%% \end{inparaenum}
This particular VAS captures the $\preceq$ partial order on
cardinality vectors as a reachability relation, as stated by the lemma
below:

\begin{lemma}\label{lemma:vas}
For any two structures $\I$ and $\I'$ sharing the same universe, we
have $\vec{t}^{\I'} \preceq \vec{t}^{\I}$ if and only if
$\vec{t}^{\I} \arrow{*}{V_{\mathcal{P}}} \vec{t}^{\I'}$.
\end{lemma}
\proof{
``$\Rightarrow$'' For any two structures $\I_1$ and $\I_2$, sharing
the same universe, we have $\vec{t}^{\I_1} \prec_1 \vec{t}^{\I_2}$ iff
there exists a set $S \subseteq \mathcal{P}$ and a predicate symbol
$p \in S$ such that: \begin{compactitem}
\item $\vec{t}^{\I_2}(S) = \vec{t}^{\I_1}(S) + 1$, 
\item $\vec{t}^{\I_2}(S \setminus \set{p}) = \vec{t}^{\I_1}(S \setminus \set{p}) - 1$, 
\item $\vec{t}^{\I_2}(T) = \vec{t}^{\I_1}(T)$, for all $T \subseteq \mathcal{P}$,
such that $T \neq S$ and $T \neq S \setminus \set{p}$.
\end{compactitem}
Then, using the fact that $\vec{t}^\I(S)
= \card{(\compt{S}{\mathcal{P}})^\I}$, for all
$S \subseteq \mathcal{P}$, we establish that
$\vec{t}^{\I_1} \prec_1 \vec{t}^{\I_2}$. Hence
$\vec{t}^{\I'} \preceq \vec{t}^{\I}$ implies the existence of a
sequence $\sigma$ of vectors from $V_{\mathcal{P}}$ such that
$\vec{t}^{\I} \arrow{\sigma}{V_{\mathcal{P}}} \vec{t}^{\I'}$.
%% Now it remains to show that there exists also an ordered such
%% sequence. To this end, we show that any pair of adjacent vectors
%% $\vec{v}(S_1,p_1)$ and $\vec{v}(S_2,p_2)$ that are not in order can be
%% swapped. In other words, if
%% $\vec{c}_1 \arrow{\vec{v}(S_1,p_1)}{} \vec{c}_2 \arrow{\vec{v}(S_2,p_2)}{} \vec{c}_3$
%% is a valid execution of $V_{\mathcal{P}}$ then there exists
%% $\vec{c}'_2$ such that
%% $\vec{c}_1 \arrow{\vec{v}(S_2,p_2)}{} \vec{c}'_2 \arrow{\vec{v}(S_1,p_1)}{} \vec{c}_3$
%% is also a valid execution of $V_{\mathcal{P}}$. Since
%% $\vec{v}(S_1,p_1)$ increases $\vec{c}_1(S_1 \setminus \set{p_1})$ by
%% one and $\vec{v}(S_2,p_2)$ decreases $\vec{c}_2(S_2)$ by one, the one
%% way $\vec{v}(S_1,p_1)$ could enable $\vec{v}(S_2,p_2)$ is when
%% $S_1 \setminus \set{p_1} = S_2$ and
%% $\vec{c}_1(S_1 \setminus \set{p_1}) = 0$. Then $\vec{v}(S_2,p_2)$
%% cannot be applied to $\vec{c}_1$ and the two vectors do not commute.
%% Assuming that $S_1 \setminus \set{p_1} = S_2$ and since $p_1 \in S_1$,
%% we have $S_2 \subset S_1$. But then $S_2 \setordneq S_1$ and the
%% vectors $\vec{v}(S_1,p_1)$ and $\vec{v}(S_2,p_2)$ are in the right
%% order, which contradicts their choice. Hence any two adjacent vectors
%% not in the right order can be swapped and the result is an ordered
%% sequence, as required.
``$\Leftarrow$'' For any two configurations $\vec{c}$ and $\vec{c}'$,
if $\vec{c} \arrow{\vec{v}}{V_{\mathcal{P}}} \vec{c}'$ for some vector
$\vec{v} \in V_{\mathcal{P}}$, then $\vec{c} \succ_1 \vec{c}'$, by
Definition \ref{def:cardinality-po}. Consequently,
$\vec{t}^{\I} \arrow{\sigma}{V_{\mathcal{P}}} \vec{t}^{\I'}$ implies
$\vec{t}^{\I'} \preceq \vec{t}^{\I}$, by straightforward induction on
the length of $\sigma$.
\qed}

%% Observe that, in an ordered execution of $V_{\mathcal{P}}$, the
%% position corresponding to a set $S \subseteq \mathcal{P}$ in all
%% configurations is first increased a number of times, possibly zero,
%% then decreased. In order to check whether an ordered sequence starting
%% with a configuration $\vec{c}\in\nat^{2^{\card{\mathcal{P}}}}$
%% corresponds to an actual execution of $V_{\mathcal{P}}$ it is thus
%% sufficient to check that the last configuration has positive entries
%% only.

For a tuple of variables $\vec{x} = \tuple{x_1, \ldots, x_k}$ and a
valuation $\nu$ mapping these variables into $\zed$, we denote by
$\nu(\vec{x})$ the tuple of integers
$\tuple{\nu(x_1), \ldots, \nu(x_k)}$. The following lemma gives an
equivalent condition for the existence of an execution in
$V_{\mathcal{P}}$, that ends in a given configuration:

\begin{lemma}\label{lemma:linsys}
Let $\vec{x}=[x_S]^T_{S \subseteq \mathcal{P}}$ and
$\vec{y}=[y_S]^T_{S \subseteq \mathcal{P}}$ be column vectors of
variables and $\set{k_{S,p} \mid S \subseteq \mathcal{P}, p \in S}$ be
variables. Then for any positive valuation $\nu$ of the variables
$\vec{x}$ and $\vec{y}$, the following are
equivalent: \begin{compactenum}
\item\label{it1:linsys} $\nu$ can be extended to a positive solution of 
the integer linear system: 
\[\vec{x} = \sum_{\scriptscriptstyle{S\subseteq\mathcal{P}, p \in S}}
k_{S,p} \cdot \vec{v}(S,p) + \vec{y}\] 
\item\label{it2:linsys}  $\nu(\vec{y}) \arrow{*}{V_{\mathcal{P}}} \nu(\vec{x})$.
\end{compactenum}
\end{lemma}
\proof{
``$\Rightarrow$'' Consider the sequence of vectors $\nu(\vec{y})
= \vec{c}_0, \vec{c}_1, \ldots, \vec{c}_k = \nu(\vec{x})$, such that
$\vec{c}_{i+1}=\vec{c}_{i}+\vec{v}(S_{i+1},p_{i+1})$ for all $1 \leq i
< k$ and the sequence of vectors
$\vec{v}(S_1,p_1), \ldots, \vec{v}(S_k,p_k)$ occur in order, each
vector $\vec{v}(S_i,p_i)$ occurring $\nu(k_{S_i,p_i}) \geq 0$ times in
the sequence. To show that this is an execution of $V_{\mathcal{P}}$,
observe that each sequence of entries
$\vec{c}_0(S), \ldots, \vec{c}_k(S)$, for some
$S \subseteq \mathcal{P}$ is first increased, then decreased, zero or
more times. Because $\vec{c}_0(S)\geq0$ and $\vec{c}_k(S)\geq0$, we
have that $\vec{c}_i(S) \geq 0$, for all $0 \leq i \leq k$. Since the
choice of $S$ was arbitrary, every vector $\vec{c}_i$ has only
positive entries, hence the sequence is an execution of
$V_{\mathcal{P}}$. ''$\Leftarrow$'' Immediate, since in every
execution $\nu(\vec{y}) \arrow{*}{V_{\mathcal{P}}} \nu(\vec{x})$, each
vector $\vec{v}(S,p)$ occurs a positive number of times and let
$\nu(k_{S,p})$ be that number. \qed}

Turning back to the original problem $\upclose{\sem{C}}
= \sem{\ppos{C}}$, we notice that the set of vectors
$\set{\vec{t}^\I \mid \exists \I' ~.~ \vec{t}^{\I'} \models
C \wedge \vec{t}^{\I'} \preceq \vec{t}^\I}$, which corresponds (up to
isomorphism) to the left-hand side of the required equality, is the
set of vectors $\nu(\vec{y})$, where $\nu$ is a positive solutions of
the linear system below:
\begin{equation}\label{eq:lhs}
\vec{x} = \sum_{\scriptscriptstyle{S\subseteq\mathcal{P}, p \in S}} k_{S,p} \cdot \vec{v}(S,p) + \vec{y}
~\wedge~ \bigwedge_{S \subseteq \mathcal{P}} \ell_S \leq \vec{x}(S) \leq u_S
\end{equation}
The formal argument combines the results of
Lemmas \ref{lemma:cardinality}, \ref{lemma:vas}
and \ref{lemma:linsys}.  Next, we show that the right-hand side
corresponds to the linear system obtained by eliminating the $\vec{x}$
and $\set{k_{S,p} \mid S \subseteq \mathcal{P}, p \in S}$ variables
from the above system.

\vspace*{\baselineskip}
\paragraph{\em Circulations in a Weighted Graph}
We eliminate the $k_{S,p}$ variables from (\ref{eq:lhs}) using
Hoffman's Circulation Theorem, given below. Let $G = (V,E)$ be a
directed graph, where $V$ is a finite set of vertices and $E \subseteq
V \times V$ a set of edges. Further, we associate each edge in $G$ a
lower and upper capacity, formally $L : E \rightarrow \nat$ and $U :
E \rightarrow \nat \cup \set{\infty}$, such that $L(e) \leq U(e)$, for
all $e \in E$. For brevity, we call $G=(V,E,L,U)$ a \emph{capacitated
graph} in the following. Given a vertex $v \in V$, we denote by
$\pre{v}$ ($\post{v}$) the set of incoming (outgoing) edges with
destination (source) $v$. We lift these notations to sets of vertices
in the usual way. A \emph{circulation} is a mapping $X :
E \rightarrow \nat$ such that, for all $v \in V$, we have
$\sum_{\text{$e \in \pre{v}$}} X(e) = \sum_{e \in \post{v}} X(e)$ and
$L(e) \leq X(e) \leq U(e)$, for all $e \in E$. The following is known
as Hoffman's Circulation Theorem \cite[Theorem 11.2]{Schrijver03}: 

\begin{theorem}\label{thm:hoffman}
Given a capacitated graph $G=(V,E,L,U)$, there exists a circulation in
$G$ if and only if $\sum_{\text{$e \in \pre{\mathcal{S}}$}}
L(e) \leq \sum_{\text{$e \in \post{\mathcal{S}}$}} U(e)$, for each set
of vertices $\mathcal{S} \subseteq V$.
\end{theorem}

We encode the existence of positive solutions of the linear integer
system (\ref{eq:lhs}) as a circulation problem in the capacitated
graph $G_{\mathcal{P}}[\vec{y}] = (2^\mathcal{P} \cup \set{\zeta},
E_{\mathcal{P}}, L_{\mathcal{P}}, U_{\mathcal{P}})$, where:
\begin{compactitem}
\item $\zeta \not \in 2^{\mathcal{P}}$ is a special vertex, not a subset of $\mathcal{P}$, 
\item $\vec{y}$ is a tuple of parameters, indexed by sets of predicate symbols,
\item for each set $S \subseteq \mathcal{P}$ there exists an edge $e=(\zeta, S)$,
with $L_{\mathcal{P}}(e)=U_{\mathcal{P}}(e)=\vec{y}(S)$,
\item for each set $S \subseteq \mathcal{P}$, there exists an edge $e=(S,\zeta)$,
with $L_{\mathcal{P}}(e)=\ell_S$ and $U_{\mathcal{P}}(e)=u_S$,
\item for each nonempty set $S \subseteq \mathcal{P}$ and each
predicate symbol $p \in S$, there exists an edge
$e=(S,S\setminus\set{p})$, with $L(e)=0$ and $U(e)=\infty$.
\end{compactitem}
Moreover, nothing else is in $E_{\mathcal{P}}$, $L_{\mathcal{P}}$ and
$U_{\mathcal{P}}$, respectively. For example, given $\mathcal{P}
= \set{a,b,c}$, the graph $G_{\mathcal{P}}$ is depicted in
Figure \ref{fig:capacitated-graph}. The following lemma relates the
existence of positive solutions of the linear integer system
(\ref{eq:lhs}) with the existence of a circulation in
$G_{\mathcal{P}}[\vec{y}]$.

\begin{figure}[htb]
\begin{center}
\scalebox{1.2}[1.25]{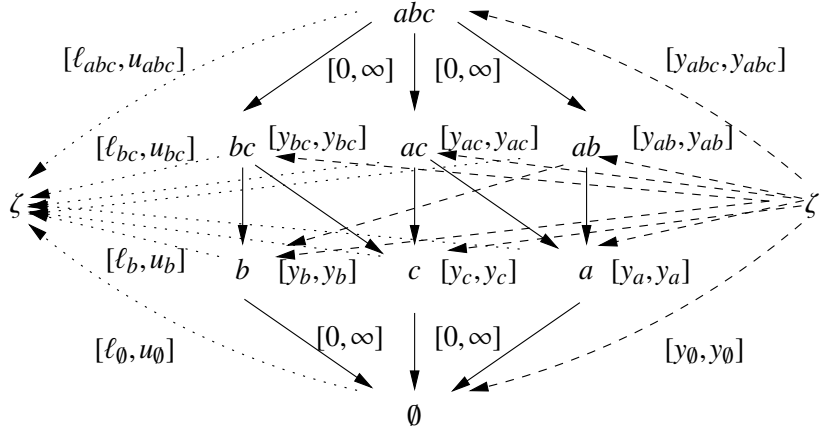}
\caption{The Capacitated Graph $G_{\set{a,b,c}}[\vec{y}]$ --- the $\zeta$ node is duplicated, for clarity.}
\label{fig:capacitated-graph}
\end{center}
\end{figure}

\begin{lemma}\label{lemma:circulation}
Given a set $\mathcal{P}$ of predicate symbols and a positive
valuation $\nu$ of the variables $\vec{y}$, the following are
equivalent:
\begin{compactenum}[(a)]
\item\label{it1:circulation} $\nu$ can be extended to a positive solution of the integer system (\ref{eq:lhs}),
\item\label{it2:circulation} the capacitated graph $G_{\mathcal{P}}[\nu(\vec{y})]$ has a circulation.
\end{compactenum}
\end{lemma}
\proof{
(\ref{it1:circulation}) $\Rightarrow$ (\ref{it2:circulation}) Assume
that $\nu$ is a positive solution of (\ref{eq:lhs}). We define the
mapping $X : E \rightarrow \nat$ as follows, for all
$S \subseteq \mathcal{P}$: \begin{compactitem}
\item $X(e) = \nu(\vec{y}_S)$ if $e=(\zeta,S)$, 
\item $X(e) = \nu(\vec{x}_S)$ if $e=(S,\zeta)$,
\item $X(e) = \nu(k_{S,p})$ if $e = (S,S\setminus\set{p})$, for some $p \in S$.
\end{compactitem}
We prove that $X$ is a circulation in $G_{\mathcal{P}}$. The condition
$L_{\mathcal{P}}(e) \leq X(e) \leq U_{\mathcal{P}}$, for all $e \in
E_{\mathcal{P}}$ is immediate, because either $L_{\mathcal{P}}(e) = 0$
and $U_{\mathcal{P}}(e) = \infty$ or it follows directly from
(\ref{eq:lhs}). It remains to check that
$\sum_{\text{$e \in \pre{u}$}} X(e) = \sum_{\text{$e \in \post{u}$}}
X(e)$, for any vertex $u \in 2^{\mathcal{P}} \cup \set{\zeta}$. If
$u=\zeta$, we have $\sum_{S \subseteq \mathcal{P}} \nu(\vec{x}(S))
= \sum_{S \subseteq \mathcal{P}} \nu(\vec{y}(S))$, because the sum of
the elements of each vector $\vec{v}(S,p)$ is zero, for any
$S \subseteq \mathcal{P}$ and $p \in S$. Else, if $u$ is some set
$S \subseteq \mathcal{P}$, we have $\nu(\vec{x}(S)) + \sum_{p \in
S} \nu(k_{S,p}) = \nu(\vec{y}(S)) + \sum{q \not\in
S} \nu(k_{S \cup \set{q},q})$.

\noindent (\ref{it2:circulation}) $\Rightarrow$ (\ref{it1:circulation})
Given a circulation $X$ in $G_{\mathcal{P}}$, we define $\nu$ as
follows, for all $S \subseteq \mathcal{P}$: \begin{compactitem}
\item $\nu(\vec{y}(S)) = X(e)$, where $e=(\zeta,S)$, 
\item $\nu(\vec{x}(S)) = X(e)$, where $e=(S,\zeta)$, 
\item $\nu(k_{S,p}) = X(e)$, where $e=(S,S\setminus\set{p})$, for some $p \in S$. 
\end{compactitem}
By definition, $\nu$ is a positive valuation. It remains to show that
$\nu$ is indeed a solution of (\ref{eq:lhs}). The condition
$\ell_S \leq \nu(\vec{x}(S)) \leq u_S$ is clearly satisfied for each
$S \subseteq \mathcal{P}$, because $L_{\mathcal{P}}(e) = \ell_S$ and
$U_{\mathcal{P}}(e) = u_S$, for each edge $e=(S,\zeta)$. To prove the
remaining condition, observe that $\nu(\vec{x}(S)) + \sum_{p \in
S} \nu(k_{S,p}) = \nu(\vec{y}(S)) + \sum_{q \not\in
S} \nu(k_{S \cup \set{q},q})$, for each $S \subseteq \mathcal{P}$,
which leads to $\nu(\vec{x}) = \sum_{S\subseteq\mathcal{P},p \in
S} \nu(k_{S,p}) \cdot \vec{v}(S,p) + \nu(\vec{y})$, as required. \qed}

Theorem \ref{thm:hoffman} gives an equivalent condition for the
existence of a circulation in $G_{\mathcal{P}}[\vec{y}]$. In the
following, we write another linear system, with unknowns $\vec{y}$
only, that captures this condition. A set of sets
$\mathcal{S} \subseteq 2^{\mathcal{P}}$ is \emph{downward closed} iff
any subset of a set in $\mathcal{S}$ is in $\mathcal{S}$. Dually,
$\mathcal{S}$ is \emph{upward closed} iff any superset of a set in
$\mathcal{S}$ is in $\mathcal{S}$. It is easy to check that the
complement of a downward (upward) closed set is upward (downward)
closed. 

Consider any capacitated graph $G_{\mathcal{P}}[\vec{y}]$, e.g. refer
to Figure \ref{fig:capacitated-graph} for an example. If
$\mathcal{S} \subseteq 2^{\mathcal{P}}$ is not downward closed, then
there exists an outgoing edge $e \in \post{\mathcal{S}}$ with $U(e)
= \infty$. Consequently, we have $\sum_{e \in \post{\mathcal{S}}} U(e)
= \infty$ in this case, thus the condition of
Theorem \ref{thm:hoffman} is trivially satisfied, for such sets. In
the light of this remark, it is obvious that we need to consider only
downward closed sets in order to characterize circulations in
$G_{\mathcal{P}}[\vec{y}]$, as in the example below:

\begin{example}\label{ex:downward-closed}
Consider the capacitated graph $G_{\set{a,b,c}}[\vec{y}]$ from
Figure \ref{fig:capacitated-graph}. The necessary and sufficient
condition for the existence of a circulation in
$G_{\set{a,b,c}}[\vec{y}]$, are partly shown below, by taking the
downward closed sets $\mathcal{S} \subseteq 2^{\mathcal{P}}$ with and
without the $\zeta$ vertex, respectively:
\[\begin{array}{rrcl}
\set{\emptyset}: & y_\emptyset & \leq & u_\emptyset \\
\set{\set{a},\emptyset}: & y_a+y_\emptyset & \leq & u_a+u_\emptyset \\
\set{\set{b},\emptyset}: & y_b+y_\emptyset & \leq & u_b+u_\emptyset \\
\set{\set{c},\emptyset}: & y_c+y_\emptyset & \leq & u_c+u_\emptyset \\
\set{\set{a},\set{b},\emptyset}: & y_a+y_b+y_\emptyset & \leq & u_a+u_b+u_\emptyset \\
\set{\set{b},\set{c},\emptyset}: & y_b+y_c+y_\emptyset & \leq & u_b+u_c+u_\emptyset \\
\set{\set{a},\set{c},\emptyset}: & y_a+y_c+y_\emptyset & \leq & u_a+u_c+u_\emptyset \\
\set{\set{a},\set{b},\set{c},\emptyset}: & y_a+y_b+y_c+y_\emptyset & \leq & u_a+u_b+u_c+u_\emptyset \\
&& \ldots
\end{array}\]
\[\begin{array}{rcl}
\set{\set{a,b,c},\set{b,c},\set{a,c},\set{a,b},\set{b},\set{c},\set{a},\emptyset}: & 
y_{abc}+y_{bc}+y_{ac}+y_{ab}+y_b+y_c+y_a+y_\emptyset & \leq \\
& u_{abc}+u_{bc}+u_{ac}+u_{ab}+u_b+u_c+u_a+u_\emptyset
\end{array}\]

\[\begin{array}{rrcl}
\set{\zeta}: & \ell_{abc}+\ell_{bc}+\ell_{ac}+\ell_{ab}+\ell_b+\ell_c+\ell_a+\ell_\emptyset
& \leq & y_{abc}+y_{bc}+y_{ac}+y_{ab}+y_b+y_c+y_a+y_\emptyset \\
\set{\zeta,\emptyset}: & \ell_{abc}+\ell_{bc}+\ell_{ac}+\ell_{ab}+\ell_b+\ell_c+\ell_a
& \leq & y_{abc}+y_{bc}+y_{ac}+y_{ab}+y_b+y_c+y_a \\
\set{\set{a},\zeta,\emptyset}: & \ell_{abc}+\ell_{bc}+\ell_{ac}+\ell_{ab}+\ell_b+\ell_c
& \leq & y_{abc}+y_{bc}+y_{ac}+y_{ab}+y_b+y_c \\
\set{\set{b},\zeta,\emptyset}: & \ell_{abc}+\ell_{bc}+\ell_{ac}+\ell_{ab}+\ell_a+\ell_c
& \leq & y_{abc}+y_{bc}+y_{ac}+y_{ab}+y_a+y_c \\
\set{\set{c},\zeta,\emptyset}: & \ell_{abc}+\ell_{bc}+\ell_{ac}+\ell_{ab}+\ell_a+\ell_b
& \leq & y_{abc}+y_{bc}+y_{ac}+y_{ab}+y_a+y_b \\
\set{\set{a},\set{b},\zeta,\emptyset}: & \ell_{abc}+\ell_{bc}+\ell_{ac}+\ell_{ab}+\ell_c
& \leq & y_{abc}+y_{bc}+y_{ac}+y_{ab}+y_c \\
\set{\set{b},\set{c},\zeta,\emptyset}: & \ell_{abc}+\ell_{bc}+\ell_{ac}+\ell_{ab}+\ell_a
& \leq & y_{abc}+y_{bc}+y_{ac}+y_{ab}+y_a \\
\set{\set{a},\set{c},\zeta,\emptyset}: & \ell_{abc}+\ell_{bc}+\ell_{ac}+\ell_{ab}+\ell_b
& \leq & y_{abc}+y_{bc}+y_{ac}+y_{ab}+y_b \\
\set{\set{a},\set{b},\set{c},\zeta,\emptyset}: & \ell_{abc}+\ell_{bc}+\ell_{ac}+\ell_{ab}
& \leq & y_{abc}+y_{bc}+y_{ac}+y_{ab} \\
&& \ldots \\
\end{array}\]
\end{example}
At this point it is easy to generalize the above example and infer an
equivalent condition for the existence of a circulation in
$G_{\mathcal{P}}[\vec{y}]$:
\begin{equation}\label{eq:circulation}
\bigwedge_{\mathcal{S} \in 2^{\mathcal{P}}} \Big(\sum_{S \in \mathcal{S}} y_s \leq \sum_{S \in \mathcal{S}} u_S
~\wedge~ \sum_{S \not\in \mathcal{S}} \ell_S \leq \sum_{S \not\in \mathcal{S}} y_S\Big)
\end{equation}

Let us now turn to the definition of $\ppos{C}$, given in terms of
complete minterms, and notice the following facts: \begin{compactenum}
\item  for each positive boolean combination $\tau \in \poslan{\phi}$, the set
$\set{S \subseteq \mathcal{P} \mid
{\compt{S}{\mathcal{P}}}^+ \rightarrow \tau}$ is upward closed and its
set of minimal elements corresponds to the minterms of $\tau$ in DNF,
\item dually, for each negative boolean combination $\tau \in \neglan{\phi}$, the set
$\set{S \subseteq \mathcal{P} \mid
{\compt{S}{\mathcal{P}}}^- \rightarrow \tau}$ is downward closed and
its set of maximal elements corresponds to the minterms of $\tau$ in
DNF,
\item because the complete minterms are pairwise disjoint, in each structure
$\I$, we have $\tau^\I
= \bigcup_{{\compt{S}{\mathcal{P}}}^+ \rightarrow \tau}
{(\compt{S}{\mathcal{P}})}^\I$, for all $\tau \in \poslan{\phi}$ and
$\tau^\I = \bigcup_{{\compt{S}{\mathcal{P}}}^- \rightarrow \tau}
{(\compt{S}{\mathcal{P}})}^\I$, for all $\tau \in \neglan{\phi}$,
\item for each positive
solution $\nu$ of (\ref{eq:circulation}), there exists a structure
$\I \in \sem{\ppos{C}}$ such that
$\nu(y_S)=\card{{(\compt{S}{\mathcal{P}})}^\I}$ and viceversa, each
structure $\I \in \sem{\ppos{C}}$ induces a positive solution of
(\ref{eq:circulation}), where
$\nu(y_S)=\card{{(\compt{S}{\mathcal{P}})}^\I}$, for all
$S \subseteq \mathcal{P}$.
\end{compactenum}
To summarize, we prove that $C \minequiv \ppos{C}$ by proving the
equivalent statement $\upclose{\sem{C}} = \sem{\ppos{C}}$. Since both
the left and the right-hand side of this equality are sets of
structures closed under isomorphism, we reduce the problem to an
equivalence between sets of integer tuples
$\set{\vec{t}^\I \mid \I \in \upclose{\sem{C}}}
= \set{\vec{t}^\I \mid \I \in \sem{\ppos{C}}}$. By
Lemma \ref{lemma:cardinality}, this is equivalent to
$\set{\vec{t}^\I \mid \exists \I' \in \sem{C}
~.~ \vec{t}^{\I'} \preceq \vec{t}^\I}
= \set{\vec{t}^\I \mid \I \in \sem{\ppos{C}}}$. Subsequently,
Lemmas \ref{lemma:vas} and \ref{lemma:linsys} prove that the left-hand
side of the latter equality is the set of positive solutions of the
linear system (\ref{eq:lhs}), restricted to the tuple of variables
$\vec{y}=\tuple{y_S}_{S \subseteq \mathcal{P}}$. By Hoffman's
Circulation Theorem (Theorem \ref{thm:hoffman}), this is the set of
positive solutions to the linear system (\ref{eq:circulation}),
obtained from the elimination of the $\vec{x}$ and $k_{S,p}$ variables
from (\ref{eq:lhs}). Finally, this set is exactly the right-hand side
of the equality above, as a result of interpreting the definition of
$\ppos{C}$ in terms of vertices of the capacitated graph
$G_{\mathcal{P}}[\vec{y}]$, on which the circulation theorem was
applied.

\fi %%% Proof of Lemma \ref{lemma:positivation} %%%

\section{Proving Deadlock Freedom of Parametric Systems}
\label{sec:verification}

\begin{figure}[htbp]
\begin{center}
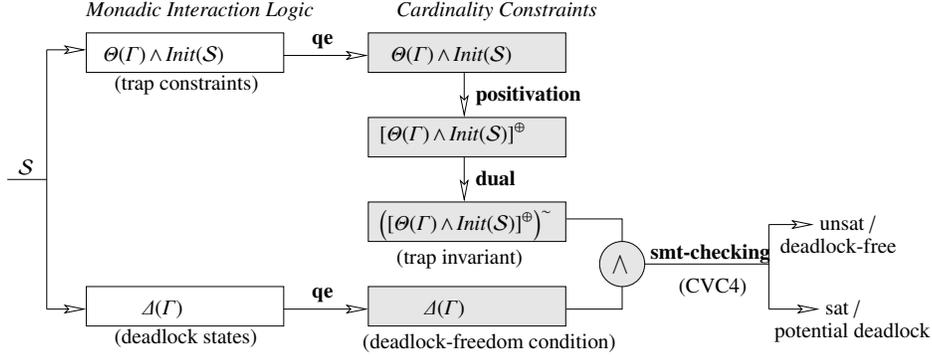
\end{center}
\caption{\label{fig:flow} Verification of Parametric Component-based Systems}
\end{figure}

We have gathered all the ingredients necessary for checking deadlock
freedom of parametric systems, using our method based on trap
invariant generation (Figure \ref{fig:flow}). In particular, we derive
a trap constraint $\trapconstraint{\interform}$ directly from the
interaction formula $\interform$, both of which are written in
\mil. Second, we compute a positive formula that preserves the set of
minimal models of $\trapconstraint{\interform} \wedge \init{\asys}$, by first converting
the \mil\ formula into a quantifier-free cardinality constraint, using
quantifier elimination, and deriving a positive \mil\ formula from the
latter. 

The conjunction between the dual of this positive formula and the
formula $\deadconstraint{\interform}$ that defines the deadlock states
is then checked for satisfiability. Formally, given a parametric
system $\asys$, with an interaction formula $\interform$ written in
the form (\ref{eq:param-interform}), the \mil\ formula characterizing
the deadlock states of the system is the following: 
\[\begin{array}{c}
\deadconstraint{\interform} \isdef \forall i_1 \ldots \forall i_\ell ~.~
\varphi \rightarrow \Big[\bigvee_{j=1}^\ell \neg\pre{p_j}(i_j) \vee 
\bigvee_{j = \ell+1}^{\ell+m} \exists i_j ~.~ \psi_j \wedge \neg\pre{p_j}(i_j)\Big]
\end{array}\]
We state a sufficient verification condition for deadlock freedom in
the parametric case:

\begin{corollary}\label{cor:param-deadlock-freedom}
  A parametric system $\asys=\tuple{\typeno{\acomptype}{1},
  \ldots, \typeno{\acomptype}{n}, \maxnf, \interform}$ is deadlock-free if 
  \[\begin{array}{c}
  \dual{\ppos{(\trapconstraint{\interform} \wedge \init{\asys})}}
  \wedge \deadconstraint{\interform} \rightarrow \false
  \end{array}\]
\end{corollary}
The satisfiability check is carried out using the conversion to
cardinality constraints via quantifier elimination \S\ref{sec:qe} and
an effective set theory solver for cardinality constraints,
implemented in the CVC4 SMT solver \cite{cvc4}. 

\section{Experimental Results}
\label{sec:experiments}

To assess our method for proving deadlock freedom of parametric
component-based system, we ran a number of experiments on systems with
a small numbers of rather simple component types, but with nontrivial
interaction patterns, given by \mil\ formulae. The {\tt task-sem}
$i/n$ examples, $i = 1,2,3$, are generalizations of the
parametric \emph{Task}-\emph{Semaphore} example depicted in
Figure \ref{fig:ex-intro-parametric}, in which $n$ \emph{Task}s
synchronize using $n$ \emph{Semaphore}s, such that $i$ \emph{Task}s
interact with a single \emph{Semaphore} at once, in a multiparty
rendez-vous. In a similar vein, the {\tt broadcast} $i/n$ examples,
$i=2,3$ are generalizations of the system in
Figure \ref{fig:bip-broadcast}, in which $i$ out of $n$ \emph{Worker}s
engage in rendez-vous on the $b$ port, whereas all the other stay idle
--- here idling is modeled as a broadcast on the $a$ ports. Finally,
in the {\tt sync} $i/n$ examples, $i=1,2,3$, we consider systems
composed of $n$ \emph{Worker}s (Figure \ref{fig:ex-intro-parametric})
such that either $i$ out of $n$ instances simultaneously interact on
the $b$ ports, or all interact on the $f$ ports. Notice that, for
$i=2,3$, these systems have a deadlock if and only if $n \neq 0 \mod
i$. This is because, if $n = m \mod i$, for some $0 < m < i$, there
will be be $m$ instances that cannot synchronize on their $b$ port, in
order to move from $w$ to $u$, in order to engage in the $f$
broadcast.

\begin{table}
{\footnotesize\begin{tabular}{| c | c | c | c | c |} 
  \hline
  example & interaction formula & t-gen & t-smt & result \\ \hline
  {\tt task-sem} 1/$n$ & $\exists i \exists j_1.~ a(i) \wedge b(j_1) ~\bigvee~ \exists i \exists j_1.~ e(i) \wedge f(j_1)$ & 22 ms & 20 ms & unsat \\ \hline
  {\tt task-sem} 2/$n$ & $\exists i \exists j_1 \exists j_2.~ j_1 \not= j_2 \wedge a(i) \wedge b(j_1) \wedge b(j_2)~\bigvee $ & & & \\ 
                 & $\exists i \exists j_1 \exists j_2.~ j_1 \not= j_2 \wedge e(i) \wedge f(j_1) \wedge f(j_2)$ & 34 ms & 40 ms & unsat \\ \hline
  {\tt task-sem} 3/$n$ & $\exists i \exists j_1 \exists j_2 \exists j_3.~ \distinct(j_1,j_2,j_3) \wedge a(i) \wedge b(j_1) \wedge b(j_2) \wedge b(j_3) ~\bigvee $ & & & \\
                 & $\exists i \exists j_1 \exists j_2 \exists j_3.~ \distinct(j_1,j_2,j_3) \wedge e(i) \wedge f(j_1) \wedge f(j_2) \wedge f(j_3)$ & 73 ms & 40 ms & unsat \\ \hline
  {\tt broadcast} 2/$n$ & $\exists i_1 \exists i_2. i_1 \not= i_2 \wedge b(i_1) \wedge b(i_2) ~\wedge $ & & & \\
                  & ~~ $\forall j.~ j\neq i_1 \wedge j \neq i_2 \rightarrow a(j) ~\bigvee \exists i. f(i)$ & 14 ms & 20 ms & unsat \\ \hline
  {\tt broadcast} 3/$n$ & $\exists i_1 \exists i_2 \exists i_3. \distinct(i_1,i_2,i_3) \wedge b(i_1) \wedge b(i_2) \wedge b(i_3) ~\wedge $ & & & \\ 
                  & ~~ $\forall j.~ j \neq i_1 \wedge j \neq i_2 \wedge j \neq i_3 \rightarrow a(j) ~\bigvee~ \exists i. f(i)$ & 409 ms & 20 ms & unsat \\ \hline 
  {\tt sync} 1/$n$ & $\exists i. b(i) ~\bigvee~ \forall i. f(i)$ & 5 ms & 20 ms & unsat \\ \hline
  {\tt sync} 2/$n$ & $\exists i_1 \exists i_2. ~i_1 \not=i_2 \wedge b(i_1) \wedge b(i_2) ~\bigvee~ \forall i. f(i)$ & 7 ms & 50 ms & sat \\ \hline
  {\tt sync} 3/$n$ & $\exists i_1 \exists i_2 \exists i_3. \distinct(i_1,i_2,i_3) \wedge b(i_1) \wedge b(i_2) \wedge b(i_3) ~\bigvee~ \forall i. f(i)$ & 11 ms & 40 ms & sat \\ \hline
\end{tabular}}
\caption{\label{tab:benchmarks} Benchmarks}
\vspace*{-2\baselineskip}
\end{table}

All experiments were carried out on a Intel(R) Xeon(R) CPU @ 2.00GHz
virtual machine with 4GB of RAM. Table \ref{tab:benchmarks} shows
separately the times needed to generate the proof obligations (trap
invariants and deadlock states) from the interaction formulae and the
times needed by CVC4 1.7 to show unsatisfiabilty or come up with a
model. All systems considered, for which deadlock freedom could not be
shown using our method, have a real deadlock scenario that manifests
only under certain modulo constraints on the number $n$ of
instances. These constraints cannot be captured by \mil\ formulae, or,
equivalently by cardinality constraints, and would require cardinality
constraints of the form $\len{t} = n \mod m$, for some constants $n,
m \in \nat$.

\section{Conclusions}
\label{sec:conclusion}

This work is part of a lasting research program on BIP linking two work
directions: \begin{inparaenum}[(1)]
\item recent work on modeling architectures using interaction
logics, and 
\item older work on verification by using invariants. 
\end{inparaenum}
Its rationale is to overcome as much as possible complexity and
undecidability issues by proposing methods which are adequate for the
verification of essential system properties.

The presented results are applicable to a large class of architectures
characterized by the \mil. A key technical result is the translation
of \mil\ formulas into cardinality constraints. This allows on the one
hand the computation of the \mil\ formula characterizing the minimal
trap invariant. On the other hand, it provides a decision procedure
for \mil, that leverages from recent advances in SMT, 
implemented in the CVC4 solver \cite{cvc4}.

Our approach sheds new light on the intricacy of the interaction structure
between components. This clearly depends on the topology of the
architecture but also on the multiplicity of interactions.  Centralized
control systems seem to be the easier to verify (parametric systems with
single controller and without interaction between components). For
distributed control systems, easier to check seem to be systems where
interactions between components are uniform – each component of a class
interacts in the same manner with all the other components.

The hardest case corresponds to systems where interaction between
components depends on a neighborhood which usually implies some arithmetic
relation between indices. To model such systems \mil\  should be extended with
arithmetic predicates on indices. This is the objective of a future work
direction.

%%%%%%%%%%%%%%%%%%%%%%%%%%%%%%%%%%%%%%%%%%%%%%%%%%%%%%%%%%%%%%%%%%%%%%%%%%%%%%%
\bibliographystyle{splncs03} \bibliography{refs}
%%%%%%%%%%%%%%%%%%%%%%%%%%%%%%%%%%%%%%%%%%%%%%%%%%%%%%%%%%%%%%%%%%%%%%%%%%%%%%%

%%%%%%%%%%%%%%%%%%%%%%%%%%%%%%%%%%%%%%%%%%%%%%%%%%%%%%%%%%%%%%%%%%%%%%%%%%%%%%%
\end{document}